\documentclass[a4paper, aps, pra, reprint, nofootinbib]{revtex4-1}
\usepackage[cm]{fullpage}
\usepackage[T1]{fontenc}
\usepackage{amssymb, amsmath, amsthm}
\makeatletter
\def\amsbb{\use@mathgroup \M@U \symAMSb}
\makeatother
\usepackage[mathscr]{euscript}
\usepackage[toc,page]{appendix}
\usepackage{mathtools}
\usepackage{verbatim}
\usepackage{bbold}

\usepackage{thmtools}
\usepackage{thm-restate}

\usepackage{hyperref}

\usepackage{cleveref}
\usepackage{xcolor}
\definecolor{darkred}{RGB}{150, 0, 0}
\definecolor{darkgreen}{RGB}{0, 100, 0}
\definecolor{darkblue}{RGB}{0, 0, 200}
\usepackage{hyperref}
\hypersetup{colorlinks, breaklinks, linkcolor=darkred, urlcolor=darkblue, citecolor=darkgreen}
\newtheorem{con}{Conjecture}

\DeclareMathOperator{\Tr}{Tr}

\newcommand{\I}{\mathbb{1}}
\newcommand{\X}{\mathsf{X}}
\newcommand{\Y}{\mathsf{Y}}
\newcommand{\Z}{\mathsf{Z}}
\def \diracspacing {0.7pt}
\newcommand{\bra}[1]{\langle #1 \hspace{\diracspacing} |} 
\newcommand{\ket}[1]{| \hspace{\diracspacing} #1 \rangle} 
\newcommand{\ave}[2][]{#1\langle #2 #1\rangle}

\newcommand{\bM}{\mathbf{M}}

\newcommand{\bW}{\mathbf{W}}

\newcommand{\R}{\mathbb{R}}
\newcommand{\C}{\mathbb{C}}

\begin{document}
\title{Extremal points of the quantum set in the CHSH scenario: conjectured analytical solution}
\author{Antoni Mikos-Nuszkiewicz, J\k{e}drzej Kaniewski}
\email{a.mikos-nuszkiewicz@uw.edu.pl, jkaniewski@fuw.edu.pl}
\affiliation{Faculty of Physics, University of Warsaw, Pasteura 5, 02-093 Warsaw, Poland}
\date{\today}
\begin{abstract}
Quantum mechanics may revolutionise many aspects of modern information processing as it promises significant  advantages in several fields such as cryptography, computing and metrology. Quantum cryptography for instance allows us to implement protocols which are device-independent, i.e.~they can be proven security under fewer assumptions. These protocols rely on using devices producing non-local statistics and ideally these statistics would correspond to extremal points of the quantum set in the probability space. However, even in the CHSH scenario (the simplest non-trivial Bell scenario) we do not have a full understanding of the extremal quantum points. In fact, there are only a couple of analytic families of such points. Our first contribution is to introduce two new families of analytical quantum extremal points by providing solutions to two new families of Bell functionals. In the second part we focus on developing an analytical criteria for extremality in the CHSH scenario. A well-known Tsirelson--Landau--Masanes criterion only applies to points with uniform marginals, but a generalisation has been suggested in a sequence of works by Satoshi Ishizaka. We combine these conditions into a standalone conjecture, explore their technical details and discuss their suitability. Based on the understanding acquired, we propose a new set of conditions with an elegant mathematical form and an intuitive physical interpretation. Finally, we verify that both sets of conditions give correct predictions on the new families of quantum extremal points.
\end{abstract}
\maketitle
%
\section{Introduction}
Entanglement between particles is one of the most fascinating phenomena in quantum mechanics and it is well-known that it can be widely used in quantum information processing. One of the most direct manifestations of entanglement is the fact that local measurements performed on entangled particles can produce correlations stronger than allowed by any classical theory. This observation was first made by John Bell in 1964 \cite{bellEinsteinPodolskyRosen1964} and the field of research focusing on such correlations is now known as Bell non-locality \cite{brunnerBellNonlocality2014}.

The importance of this phenomenon comes not just from its fundamental meaning, but also due to the possibility of real-world applications. The existence of quantum cryptographic protocols, which are device-independent (e.g.~to generate certifiable secure randomness \cite{herrero-collantesQuantumRandomNumber2017} or secure cryptographic key \cite{pironioDeviceindependentQuantumKey2009, boaronSecureQuantumKey2018}) is inherently linked to Bell non-locality. More specifically, the security of device-independent protocols is based only on the observed non-local statistics and the assumption that the device is governed by quantum mechanics.

In a bipartite Bell scenario we have two devices (following the standard convention, we call the involved parties Alice and Bob), and each of them can choose one out of $n$ possible measurement settings and produce one out of $m$ possible outcomes. In such a scenario, a complete characterisation of the input-output behaviour corresponds to a so-called probability point. Points obtained from all possible quantum devices form a convex set called the quantum set. If we restrict our attention to classical devices, we obtain another convex set called the local set. It is well-known that the local set is a polytope and that it is a proper subset of the quantum set \cite{brunnerBellNonlocality2014}. Ideally, we would like to have an analytic criterion determining whether a given probability point belongs to the quantum set. However, this turns out to be a hard task even in the simplest non-trivial Bell scenario called the CHSH scenario which corresponds to $n=m=2$. On the other hand, it is known that in the CHSH scenario all the extremal points can be achieved using a two-qubit realisation, which makes the search for extremal points slightly more tractable. Therefore, in this work we focus on looking for an analytic condition to determine whether a specific two-qubit realisation in the CHSH scenario produces an extremal quantum point. Since in the CHSH scenario the
quantum set equals the convex hull of its extremal points (an immediate consequence of the Krein--Millman theorem), identifying its extremal points is morally almost as good as determining the actual quantum set.

Let us briefly review the existing literature on the topic. For quantum points with zero marginals, i.e.~where the local outcomes of Alice and Bob are uniformly random, a condition for extremality can be deduced from the well-known Tsirelson--Landau--Masanes condition \cite{tsirelsonQuantumAnaloguesBell1987,landauEmpiricalTwopointCorrelation1988,masanesNecessarySufficientCondition2003}. In fact, a recent paper by Le et al.~provides a detailed overview of the different formulations of the problem and carefully describes that part of the boundary of the quantum set \cite{leQuantumCorrelationsMinimal2022}. Much less is known about points where the marginals are non-zero. There, we know some families and isolated points \cite{yangRobustSelfTesting2013, bampsSumofsquaresDecompositionsFamily2015, wagnerDeviceindependentCharacterizationQuantum2020, hardyQuantumMechanicsLocal1992, raiDeviceindependentBoundsCabello2021}  which are provably extremal, but a complete characterisation of the extremal points is not known. There exists a hypothetical set of analytic conditions proposed in a sequence of works by Ishizaka \cite{ishizakaCryptographicQuantumBound2017, ishizakaNecessarySufficientCriterion2018, ishizakaGeometricalSelftestingPartially2020}, but these conditions do not seem to have an intuitive physical interpretation and, moreover, there is no proof that they are correct. Finally, this hypothesis has only been tested on a small set of extremal points.
%

The first contribution of this work is the discovery of new quantum extremal points with non-zero marginals. More specifically, we derive two new families of provably extremal points. The second contribution is a thorough analysis of the conjecture of Ishizaka. We have gone through all the arguments, we have rephrased some of them as stand-alone lemmas and we have identified some subtle issues with the formulation. Remarkably, we find that this hypothesis gives correct predictions for all the families we have derived. Finally, we use the understanding gained in the process to propose a new set of analytic conditions for extremality. Our conditions are mathematically elegant and can be given a physical interpretation. We have checked that, similar to the hypothesis of Ishizaka, our hypothesis gives the correct predictions for all the points we have investigated.
%
\section{Preliminaries}
In this section we provide some basic knowledge about the quantum set in the CHSH scenario and then present a few analytic and numerical methods that we use to analyse it. We assume that the reader is familiar with the basic notions of Bell nonlocality.\footnote{For a gentle introduction to nonlocality we recommend the lecture notes of Scarani~\cite{scaraniDeviceindependentOutlookQuantum2015} or the review paper by Brunner et al.~\cite{brunnerBellNonlocality2014}. More details on the CHSH scenario can be found in Ref.~\cite{gohGeometrySetQuantum2018}.}
\subsection{Extremal points in the CHSH scenario}
\label{SecExtremal}
Our goal is to find the simplest characterisation of the extremal points of the quantum set in the CHSH scenario. Note that we only care about nonlocal extremal points (the only local extremal points are the deterministic points) and in the CHSH scenario being nonlocal is equivalent to violating one of the CHSH inequalities.

It is well-known that every quantum point can be achieved by performing projective measurements on a pure quantum state. What is special about the CHSH scenario is that all the extremal points have a quantum realisation based on a two-qubit state \cite{masanesExtremalQuantumCorrelations2005}. From now on, we will call a quantum point a two-qubit point if it can be obtained by measuring a pure two-qubit state using one-rank measurements. In the case of two-outcome measurements, the measurements can be equivalently expressed as observables where the two outcomes are associated to values $\pm 1$. This gives rise to Hermitian operators whose spectrum is contained in the interval $[-1, 1]$. We will denote the observables of Alice and Bob by $A_x$ and $B_y$, where $x,y \in \{0,1\}$ are the measurement settings, and note that observables arising from projective measurements satisfy $A_{x}^{2} = \I$ and $B_{y}^{2} = \I$.
In Appendix~\ref{AppParam} we show that every extremal point in the CHSH scenario has a realisation of the following form:
%
\begin{align}
\ket{\psi} &= \cos \tfrac{\theta}{2} \ket{00} + \sin \tfrac{\theta}{2} \ket{11},  \label{paramI}\\
A_x &= \cos a_x \Z + \sin a_x \X, \\
B_y &= \cos b_y \Z + \sin b_y \X, \label{paramII}
\end{align}
where $\X$ and $\Z$ are the Pauli matrices, $a_x, b_y \in [0,2\pi)$ and $\theta \in [0, \frac{\pi}{2}]$. We refer to this form as the canonical two-qubit parametrisation or realisation.

The fact that we can restrict our attention to states coming from this 1-parameter family is a direct consequence of the Schmidt decomposition and the invariance of probabilities under local unitaries (applied to both the state and the observables). The fact that we can further assume all the observables to lie in the $\X\Z$ plane is often considered folklore knowledge, but is actually non-trivial. 
Using this parametrisation, the probability point can be uniquely expressed in terms of eight components: $\mathbf{P} = \big( \langle A_x \rangle,  \langle B_y \rangle, \langle A_x B_y\rangle \big)$:
\begin{align}
\langle A_x \rangle &= \cos \theta \cos a_x,\label{real1}\\
\langle B_y \rangle &= \cos \theta \cos b_y,\\
\langle A_x B_y \rangle &= \cos a_x \cos b_y +  \sin \theta \sin a_x \sin b_y.\label{real2}
\end{align}
Here, $\langle A_x \rangle$ and $\langle B_y\rangle$ are called the marginals, while $\langle A_x B_y\rangle$ are called the correlators. This parametrisation contains all the extremal points, but it also contains some non-extremal points. The main goal of this work is to find analytic criteria which would allow us to tell the two classes apart.

Since we are interested in the simplest parametrisation that still contains all the extremal quantum points, let us make two additional comments:
\begin{itemize}
    \item Changing all the observable angles according to the transformation $a_{x} \mapsto - a_{x}$ and $b_{y} \mapsto - b_{y}$ leaves the statistics unchanged (the $\cos a_{x}, \cos b_{y}$ terms remain unchanged, while $\sin a_{x}, \sin b_{y}$ are multiplied by $-1$, which cancels out). This means that we may restrict the angle of one of the observables to half of its original domain.
    \item In the special case of the maximally entangled state, i.e.~$\theta = \pi/2$, it is easy to check that
\begin{align}
\langle A_x \rangle &= \langle B_y \rangle = 0,\\
\langle A_x B_y \rangle &= \cos ( a_x - b_y ).
\end{align}
Since the marginals vanish and the correlators only depend on the difference between the angles, we may without loss of generality choose one of the angles to vanish, e.g.~$a_{0} = 0$.
\end{itemize}
\subsection{Local and non-signalling sets}
The local set is the set of points that can be obtained from a local hidden-variable model. In the CHSH scenario it is a polytope with $16$ vertices, known as the deterministic points, that have the following structure: $P_D = (c_0,c_1,d_0,d_1,c_0d_0, c_0d_1,c_1d_0,c_1d_1 )$  where $c_x, d_y = \pm 1$.  

The non-signalling set is the largest set of correlations, which does not violate the fundamental principle that measurements on space-like separated systems should not allow Alice and Bob to exchange information. This set also turns out to be a polytope and in the CHSH scenario its vertices consist of the deterministic points and $8$ additional points known as the PR boxes. The PR boxes are the points of the form $(0,0,0,0,e_1,e_2,e_3,e_4)$ where $e_1,e_2,e_3,e_4 \in \{-1,1\}$ and $e_1 e_2 e_3 e_4 = -1$.
\subsection{Local, quantum and non-signalling values} \label{setValue}
Since a probability point can be thought of as a vector, $P \in \R^8$, we can define linear functionals that act on these vectors and we will denote them by $F \in \R^8$. For every such a functional we can define the local, quantum and non-signalling values as follows:
\begin{align}
\beta_S := \max_{P\in S} F \cdot P,
\end{align}
where $S$ denotes a corresponding set of points. Since all three sets are convex and compact, it suffices to take the maximum over extremal points. In the case of the local (non-signalling) value the problem reduces to maximising over the deterministic points (the deterministic points and PR boxes). We are interested in functionals whose quantum value is strictly larger than the local value, $\beta_{Q} > \beta_{L}$. When this happens, we say that the functional exhibits a quantum advantage and the corresponding point $P$ is non-local.
\subsection{Non-negativity facets}
In general, the probability point is described by probabilities $p(ab|xy)$, which tell us what is the probability, that Alice measures output $a$ and Bob measures output $b$, when they choose corresponding measurements $x$ and $y$. Our parametrisation is simplified, but when we are given a point $P= \big( \langle A_x \rangle,  \langle B_y \rangle, \langle A_x B_y\rangle \big)$ we can find all probabilities $p(ab|xy)$ using the following formula
\begin{align}
    p(ab|xy) &= \frac{1}{4}\big(1 + (-1)^{a}\ave{A_x} + (-1)^{b}\ave{B_y} +\nonumber\\
    &(-1)^{a+b}\ave{A_x B_y} \big). \label{prob}
\end{align}

We know, that all probabilities $p(ab|xy)$ have to lie in the interval $[0,1]$. It is easy to see, that the upper bound is always satisfied when the absolute values of all the marginals and correlators are bounded by $1$. Therefore the only non-trivial condition is that $p(ab|xy)\geq 0$ for all $x,y,a,b$. In the parametrisation from Eqs.~\eqref{real1}--\eqref{real2} it can be written as follows:
\begin{align}
    1 + (-1)^{a}\ave{A_x} + (-1)^{b}\ave{B_y} +(-1)^{a+b}\ave{A_x B_y}\label{prob2} \geq 0
\end{align}
Each of these inequalities defines a facet of the non-signaling set (which are also facets of quantum and local sets). We call them non-negativity facets.
\subsection{Methods for investigating extremal points} \label{method1}
In this section, we introduce a few methods which turn out to be useful for numerical and analytical analysis of the quantum extremal points.
\subsubsection{Maximising a functional} \label{methodI}
The aim of this method is to characterise the set of maximisers for a particular functional. If the maximiser turns out to be unique, then we have identified an exposed point of the quantum set. Since all the extremal points have two-qubit realisations in the canonical form, it is sufficient to optimise over two-qubit points. We can define the problem as follows: for a given functional $F$, find a two-qubit point that maximises the following quantity
\begin{align}
\beta_{\max} = \max_{P\in Q_2} F \cdot P ,
\end{align}
where $Q_2$ denotes the set of two-qubit quantum points in the canonical form.

Let us  now define the Bell operator $\bW$ for a given functional $F$ and probability point $P$:
\begin{align}
\bW &= F_1 A_0\otimes\I + F_2 A_1\otimes\I + F_3\I\otimes B_0+ F_4\I\otimes B_1 +\nonumber\\
&+ F_5 A_0\otimes B_0+ F_6 A_0\otimes B_1+ F_7 A_1\otimes B_0+ F_8 A_1\otimes B_1 \label{BellOperator}
\end{align}
It is clear that $\beta_{\max}$ is the expected value of the Bell operator acting on point $P$:
\begin{align}
\beta_{\max} = \max_{A_x, B_y,\rho} \Tr(\bW \rho).
\end{align}
We can easily perform the maximisation over the state $\rho$, because it is equivalent to taking the largest eigenvalue of the Bell operator
\begin{align}
\beta_{\max} = \max_{A_x, B_y} \lambda_{\max}(\bW)
\end{align}
and the corresponding eigenvector is the optimal quantum state. Since we do not need an explicit expression for the state, we can introduce an alternative parametrisation of two-qubit points, different from the parametrisation given in Eqs.~\eqref{paramI}--\eqref{paramII}, which has fewer free parameters. As we do not care about the choice of local bases, the only relevant parameters are angles between observables $A_0, A_1$ and $B_0, B_1$ (in the Bloch sphere picture). Hence we can set the $A_0$ and $B_0$ in the direction of $\Z$ axis and then rotate $A_1, B_1$ the $\X\Z$ plane:
\begin{align}
A_0 &= \Z,\label{SecondParam1}\\
A_1 &= \cos a \Z + \sin a \X,\\
B_0 &= \Z,\\
B_1 &= \cos b \Z + \sin b \X. \label{SecondParam2}
\end{align}
for $a,b \in [0,\pi]$. Then we can write:
\begin{align}
\beta_{\max} = \max_{a,b \in [0,\pi]} \lambda_{\max} \big(\bW(a,b)\big) \label{Imax}
\end{align}
so we can see, that the whole problem is simplified to an optimisation over two real parameters $a,b$.

Let us observe, that if $a$ (or $b$) is equal to the extremal values $0, \pi$, then the observables of Alice (or Bob) commute, which means, that the resulting statistics are necessarily local. Therefore, if there exists a global maximum for $a,b \in (0, \pi)$, then the resulting point is a non-local extremal point of the quantum set.
\subsubsection{Numerical method: finding an exposing functional} \label{methodIII}
In this method, we are given a specific two-qubit canonical point and the goal is to determine whether the resulting statistics correspond to an exposed point of the quantum set. If there exists a functional that is maximised by this point and there is no other point that reaches the same value, then we have a proof, that the given point is exposed. We can solve this problem using so-called linear programming.

We define our linear program (LP) as follows: For a given canonical two-qubit quantum point $P \in \R^8$ and its realisation~\eqref{paramI}--\eqref{paramII} we are looking for a functional $F$ that maximises the following quantity:
\begin{align}
I_{\max} = \max_{F\in \R^8} F\cdot P
\end{align}
and obeys the following constraints
\begin{align}
F \cdot \frac{\partial P}{\partial \theta} &= 0, \quad F \cdot \frac{\partial P}{\partial a_0} = 0, \quad  F \cdot \frac{\partial P}{\partial a_1} = 0, \\
\quad F \cdot \frac{\partial P}{\partial b_0} &= 0, \quad  F \cdot \frac{\partial P}{\partial b_1} = 0, \\
 F \cdot P_j &\leq 1 \quad j = 1 \cdots 16. \label{norm}
\end{align}
where $P_j$ are deterministic points. The first five conditions ensure that the probability point $P$ is at least a local maximiser for the functional $F$. The last condition corresponds to some arbitrary normalisation. Note that this method can only give us a candidate for an exposing functional, because it does not rule out the existence of another probability point that also maximises this functional. Nevertheless, it is useful for numerical calculations because linear programs can be solved efficiently. Let us note that if a given non-local point is exposed, then the value of the LP must be greater than one $I_\text{max}>1$: an exposed quantum point must exhibit some quantum advantage (the local value is fixed to be one). Similarly, when the point lies on the boundary of the quantum set, then the value resulting from the LP must satisfy $I_\text{max} \geq 1$. We can deduce it from the fact that a boundary point always maximises a certain functional (not necessarily unique). Therefore, we can use this functional in the linear program and scale it until its local value reaches $1$. This implies that if the LP produces $I_\text{max} <1$, then the probability point $P$ must lie in the interior of the quantum set.

Note that by combining the methods explained in Sections~\ref{methodI} and \ref{methodIII}, we can find a certificate that a given probability point is exposed. Firstly, for a given point we use method \ref{methodIII} in order to find a candidate for the exposing functional. Then we use method \ref{methodI} to find the optimal probability point for this functional. If one obtains the same point as the initial one and it is a unique maximiser, then we are guaranteed that this point is exposed. Unfortunately, this method does not always lead to clear conclusions. If the resulting point is the same as the initial one, then we know that it is an extremal point, but if not, then there are two options. First is that it is not an extremal point. The second case is that the method \ref{methodIII} returned a functional that is also maximised by another point, but there may exist another functional that is maximised only by the specified probability point. In the second case, we do not obtain the answer if this is an extremal point, but in practice, this case occurs very rarely. 

We also tried to add more constraints to this linear program.  Two new conditions follow from the fact that  the state is the eigenstate of the Bell operator. However, numerical calculation suggests that these conditions are linearly dependent on the rest of the conditions, so they do not affect the result. 
\section{Results}
\subsection{New families of the quantum extremal points} \label{families}
In this section, we introduce three new families of quantum points, which we analytically prove to be extremal. 
\subsubsection{Double-Tilted CHSH functionals}\label{doubleTilted}
Let us consider a two-parameter family of functionals, which is a generalisation of the tilted-CHSH functional \cite{acinRandomnessVsNon2012}:
\begin{align}
F = \big( \alpha \cos(\tfrac{\phi}{2})  ,\alpha \sin(\tfrac{\phi}{2}) , 0,0,1,1,1,-1 \big), \label{functional1}
\end{align}
for $\alpha \in \R,  \phi \in [0, 4 \pi)$.  The corresponding Bell operator reads:
\begin{align}
\bW &= \alpha \cos(\tfrac{\phi}{2}) A_0 \otimes \I +\alpha \sin(\tfrac{\phi}{2}) A_1 \otimes \I + A_0 \otimes B_0 \nonumber\\
&+ A_0 \otimes B_1 + A_1 \otimes B_0 - A_1 \otimes B_1.   \label{bellop2}
\end{align}
We can consider a few local transformations of observables and their combinations: $A_0 \leftrightarrow A_1, \quad  B_1 \rightarrow -B_1$,  $A_1 \rightarrow -A_1,  \quad  B_0 \leftrightarrow B_1$, $A_0 \rightarrow -A_0,  \quad  B_0 \leftrightarrow -B_1$. If we apply them to the Bell operator, then we obtain an equivalent Bell operator and hence, we see that it is sufficient to analyse functionals satisfying $\alpha \geq 0$ and $\phi \in [0, \frac{\pi}{2}]$. To further narrow down the range of values for which the quantum value might be larger than the classical value, let us first compare the local and no-signalling values of this functional. 

Using definitions of the local and non-signalling values from section \ref{setValue} it is easy to compute that:
\begin{align}
\beta_L &=  \alpha \cos(\frac{\phi}{2}) + \alpha \sin(\frac{\phi}{2}) + 2,\\
\beta_{NS} &= \max(\beta_L, 4).
\end{align}
Quantum advantage is only possible when $\beta_{NS} > \beta_L$, which requires that $\beta_L < 4$. Since $\cos (\frac{\pi}{2}) + \sin (\frac{\pi}{2}) \geq 1$ for $\phi \in [0, \frac{\pi}{2}]$, this implies that $\alpha <2$. Therefore, from now on, let us restrict our attention to the range $\alpha \in [0,2)$ and $\phi \in [0, \frac{\pi}{2}]$. In the following derivation, we will apply the method given in Section~\ref{methodI} to analytically compute the quantum value of the given functional and find a realisation that reaches it.

We begin with constructing the Bell operator using the parametrisation given in Eqs.~\eqref{SecondParam1}--\eqref{SecondParam2}. We obtain a $4\times4$ matrix whose eigenvalues can be computed analytically and they are properly defined in the considered ranges of parameters (see Appendix~\ref{AppEigen} for details). The eigenvalues are given by:
\begin{widetext}
\begin{align}
\begin{cases}
\lambda_{1} &= \Big( 4 + \alpha^2 + \alpha^2 \cos (2a) \sin \phi + 2\big(1 + 3 \alpha^2 - \cos(2 b) + \alpha^2 \cos b \cos \phi \\
& + 4 \alpha^2 \cos (2a) \sin \phi +\cos(4 a) (-1 + \alpha^2 + \cos(2 b) - \alpha^2 \cos b \cos \phi ) \big)^{\frac{1}{2}} \Big)^{\frac{1}{2}} \\
\lambda_{2} &= \Big( 4 + \alpha^2 + \alpha^2 \cos (2a) \sin \phi - 2\big(1 + 3 \alpha^2 - \cos(2 b) + \alpha^2 \cos b \cos \phi \\
& + 4 \alpha^2 \cos (2a) \sin \phi + \cos(4 a) (-1 + \alpha^2 + \cos(2 b) - \alpha^2 \cos b \cos \phi) \big)^{\frac{1}{2}} \Big)^{\frac{1}{2}}\\
\lambda_{3} &= -\lambda_{1}\\
\lambda_{4} &= -\lambda_{2}
\end{cases}\label{EigenMax}
\end{align}
\end{widetext}

Since square roots are non-negative, $\lambda_1$ is the largest eigenvalue. Now, we would like to find the angles $a$ and $b$ that maximise $\lambda_{1}$.
\paragraph{Optimising over $b$}
It turns out that finding the optimal angle $b$ can be done analytically and that the result does not depend on the angle $a$. In order to do so, it suffices to maximise the only part of $\lambda_1$ that depends on $b$, namely:
\begin{equation}
\begin{aligned}
f(b) &:=   - \cos(2 b) + \alpha^2 \cos b \cos \phi \\
&+\cos(4 a) ( \cos(2 b) - \alpha^2 \cos b \cos \phi  ). 
\end{aligned}
\end{equation}
Now, let us treat $\cos b$ as a new variable in this function:

\begin{align}
\Tilde{f}(x) &:=   - (2 x^2 - 1) + \alpha^2 x \cos \phi \nonumber\\
&+\cos(4 a) ( (2 x^2 - 1) - \alpha^2 x \cos \phi  )\\
&= (1 - \cos(4a))\big( -2x^2 + \alpha^2 \cos\phi x + 1 \big).
\end{align}

This is a quadratic function of $x$ and its maximum is reached for:
\begin{align}
x_{\text{max}} = \frac{\alpha^2 \cos \phi}{4}.\label{optB}
\end{align}
Thanks to the condition $\alpha < 2$ the right-hand side belongs to the interval $[0, 1)$, so we can always choose an angle $b_\text{opt}$ such that $\cos b_\text{opt} = x_\text{max}$.
\paragraph{Optimising over $a$}
Now, we can plug $\cos(b_{\text{opt}})$ into the Eq.~\eqref{EigenMax} and then define a function $g(a)$ that represent the part of the Eq.~\eqref{EigenMax} that depends on $a$:
\begin{widetext}
\begin{equation}
\begin{aligned}
g(a) &:=  \alpha^2 \cos (2a) \sin \phi + 2\big(1 + 3 \alpha^2 - \cos(2 b_\text{opt}) + \alpha^2 \cos b_\text{opt} \cos \phi +\cos(4 a) (-1 + \alpha^2 + \cos(2 b_\text{opt}) \\ 
&-\alpha^2 \cos b_\text{opt} \cos \phi + 4 \alpha^2 \cos (2a) \sin \phi ) \big)^{\frac{1}{2}}.  
\end{aligned}
\end{equation}
As before, we introduce a new function $\Tilde{g}$, where  $y:=\cos (2a)$ is a new variable:
\begin{equation}
\begin{aligned}
\Tilde{g}(y) &:=  \alpha^2  y \sin \phi  + 2\big(1 + 3 \alpha^2 - \cos(2 b_\text{opt}) + \alpha^2 \cos b_\text{opt} \cos \phi +(2y^2-1) (-1 + \alpha^2 + \cos(2 b_\text{opt}) \\ 
&-\alpha^2 \cos b_\text{opt} \cos \phi + 4 \alpha^2 y\sin \phi   ) \big)^{\frac{1}{2}}.  
\end{aligned}
\end{equation}

\end{widetext}
 Hence, if we compute the derivative of $\Tilde{g}(y)$ and set it to $0$, we obtain a quadratic equation for $y$, which leads to the following solutions:
\begin{align}
y_1 &= \frac{\alpha^2\sin\phi}{4-\alpha^2},\\
y_2 &= \frac{\alpha^2 \sin \phi \big(96 - 16\alpha^2 - \alpha^4(1+\cos(2\phi))\big)}{(4-\alpha^2)\big(32 - 16\alpha^2 + \alpha^4(1+ \cos(2\phi))\big)}. \label{optA}
\end{align}
We want $y_{1,2}= \cos(2a)$, hence these two solutions correspond to a maximum and minimum of the quadratic equation when $y_1,y_2 \in [-1,1]$. If $y$ corresponding to the maximum is out of this interval, then the maximal value is reached for $a\in \{ 0, \pi \}$, which gives a local value. 
\paragraph{Quantum value}
To find out which solution $y_1, y_2$ corresponds to the maximal eigenvalue, we plug both of them into Eq.~\eqref{EigenMax}. The solution that gives a larger value is the one that we are looking for:
\begin{align}
\beta_Q^1 &= \sqrt{\frac{32  - \alpha^4(1+ \cos(2\phi))}{(4 - \alpha^2)}},\\
\beta_Q^2 &= \sqrt{2}\sqrt{\frac{(4 - \alpha^2)\big(32  - \alpha^4(1+ \cos(2\phi))\big)}{32 - 16\alpha^2 + \alpha^4(1+ \cos(2\phi))}}. \label{quantumValue}
\end{align}
The value under the square root in $\beta_Q^2$ can be negative for some values of $\alpha, \phi$. The reason is that now, we have an additional assumption that $|y_2|\leq 1$. However, we checked that the initial eigenvalue $\lambda_1$ is properly defined for all values of its parameters. Hence, when this assumption is satisfied, then $\beta_Q^2$ is properly defined. Comparing $\beta_Q^1$ and $\beta_Q^2$ with algebraic transformations leads to the simplified inequality:
\begin{align}
\beta_Q^1 \leq \beta_Q^2 \iff 0 \leq\alpha^4 \sin^2\phi.
\end{align}
The equality is achieved only when $\alpha=0$ or $\phi = 0$. Hence, the angle $a$ that may give us a quantum advantage, corresponds to $y_2$. Therefore, if $|y_2|\leq 1$ then $a_{\text{opt}}= y_2$ and the quantum value equals $\beta_Q^2$. If $a_\text{opt}, b_\text{opt} \not\in \{0,\pi\}$ and $\beta_Q^2 > \beta_L$, then the functional exhibits the quantum advantage. In this case, one can check numerically that $\beta_Q^2 > \beta_L$ when all other conditions are satisfied.

\paragraph{Quantum extremal points}
Every functional satisfying the conditions $|y_2|\leq 1$,  $a_\text{opt}, b_\text{opt} \not\in \{0,\pi\}$ corresponds to a quantum extremal point. This  region is plotted in Fig.~\ref{regionTilted}. While we have found analytic expressions for Alice and Bob's optimal angles, computing an analytical form of the optimal state is difficult. Nevertheless, with these analytical expressions the optimal state and, hence, the entire probability point can be efficiently computed numerically. 

\subsubsection{Generalised Wolfe--Yelin functionals} \label{generalWolf}
Let us consider a 2-parameter family of Bell functionals, which generalise a functional introduced in Ref.~\cite{wolfeNewQuantumBounds2012}.
\begin{align}
F = (\alpha_0, \alpha_0, \alpha_1, 0, 1, 1, 1, -1), \label{functional}
\end{align}
where parameters  $\alpha_0 \in (-1,1), \alpha_1 \in [0,2]$. This functional corresponds to the following Bell operator:
\begin{align}
\bW &= \alpha_0 ( A_0 \otimes \I + A_1 \otimes \I) + \alpha_1 \I \otimes B_0 + \nonumber\\
&+ A_0 \otimes B_0 + A_0 \otimes B_1 + A_1 \otimes B_0 - A_1 \otimes B_1. \label{bellOperator}
\end{align}
Using definitions from Section~\ref{setValue}, we get local and non-signalling values:
\begin{align}
\beta_{L} &= \max\{ 2  \alpha_0  + \alpha_1  + 2, -2  \alpha_0  - \alpha_1  + 2, \alpha_1 + 2 \}, \label{localValue}\\
\beta_{NS} &= \max \{ \beta_L, 4\}.
\end{align}
The range of $\alpha_0, \alpha_1$ results from numerical calculations showing that quantum advantage is not possible outside this region and from symmetries discussed in the case of the Double-Tilted CHSH functionals family. 
\paragraph{Bell Operator}
For these functionals it is convenient to adopt the following parametrisation of observables:
\begin{align}
A_0 &= \cos(\frac{a}{2}) \Z + \sin(\frac{a}{2}) \X,\label{realistaionI}\\ 
A_1 &= \cos(\frac{a}{2}) \Z - \sin(\frac{a}{2}) \X,\\
B_0 &=  \Z, \\
B_1 &= \cos(b) \Z + \sin(b) \X,\label{realistaionII}
\end{align}
where $a,b \in [0,\pi]$. This is clearly equivalent (up to a local unitary on Alice) to the parametrisation given in Eq.~\eqref{SecondParam1}--\eqref{SecondParam2}), and, as we will see later, it ensures that the Bell operator takes a simple block-diagonal form, which makes it easier to compute the eigenvalues.
Writing down the Bell operator $\mathbf{W}$ in the computational basis gives:
\begin{widetext}
\begin{align}
\mathbf{W} = 
\begin{pmatrix}
\alpha_1+2\cos(\frac{a}{2})(1+\alpha_0) & 0 & 2\cos b \sin(\frac{a}{2}) &  2\sin b \sin(\frac{a}{2})\\
0 & -\alpha_1+2\cos(\frac{a}{2})(-1+\alpha_0) &  2\sin b \sin(\frac{a}{2}) & -2\cos b \sin(\frac{a}{2})\\
2\cos b \sin(\frac{a}{2}) &  2\sin b \sin(\frac{a}{2}) & \alpha_1-2\cos(\frac{a}{2})(1+\alpha_0) & 0\\
2\sin b \sin(\frac{a}{2}) & -2\cos b \sin(\frac{a}{2}) & 0 & -\alpha_1+2\cos(\frac{a}{2})(1-\alpha_0)\\
\end{pmatrix}. \label{wolfoperator}
\end{align}
\end{widetext}
Then, our task is to find the maximal eigenvalue of this matrix, optimising it over $a,b$ for fixed $\alpha_0, \alpha_1$. 
\paragraph{Eigenvalues and optimising over $b$}
The Bell operator in the computational basis exhibits an interesting structure. If we split it into $2 \times 2$ blocks, then we see that the blocks on the main diagonal are already diagonal, while the off-diagonal blocks are proportional to a unitary. Moreover, the only dependence on $b$ appears in the off-diagonal blocks. This structure allows us to show that if such a functional exhibits a classical-quantum gap, then the quantum value is achieved for $b = \frac{\pi}{2}$ and the following condition has to obey $2\cos(\frac{a}{2}) \geq  \alpha_1$(see Appendix~\ref{Apppi2} for details). Once we set $b = \frac{\pi}{2}$, then the problem reduces to diagonalising $2 \times 2$ matrices and the maximal eigenvalue of the Bell operator equals:
\begin{align}
\lambda &= 2 \cos (\frac{a}{2}) + \Big( 2 + \alpha_1^2 + 2 \alpha_0^2 \nonumber\\
&+ 4 \alpha_0\alpha_1 \cos (\frac{a}{2}) - 2 \cos a(1 -  \alpha_0^2)\Big)^{\frac{1}{2}}.\label{maxEigenvalue}
\end{align}
The value under the square root in the equation above is non-negative for all $\alpha_0, \alpha_1, a$ that we consider. One can simply check it because it is a quadratic function in terms of $\cos(\frac{a}{2})$.
\paragraph{Optimising over $a$}
Now, the problem of finding the largest eigenvalue of $\bW$ is reduced to maximising $\lambda$ over $a$. In order to do that, we can proceed as in the case of the Double-Tilted CHSH functional family and introduce a new variable $x = \cos(\frac{a}{2})$ and treat $\lambda(x)$ as a function of $x$ and set its derivative to $0$. This procedure leads to the quadratic equation for $x$:
\begin{align}
&4(-\alpha_0^4 + 3\alpha_0^2- 2)x^2 +4\alpha_1 \alpha_0(2-\alpha_0^2)x \nonumber\\
&+ 4 + \alpha_1^2(1 -\alpha_0^2) = 0. \label{quadratic2}
\end{align} 
We would like to study how many solutions it has because they are the candidates for a value of $a$ that gives a maximal eigenvalue. In order to do that, let us compute the discriminant:
\begin{align}
\Delta &= 16(2-\alpha_0^2)(\alpha_1^2 + 4(1-\alpha_0^2)).
\end{align}
For the range of $\alpha_{0}, \alpha_{1}$ that we are interested in, this is always greater than $0$, and hence, there are two distinct solutions. however, we do not know yet if they belong to the appropriate range $[-1,1]$ otherwise, we cannot interpret them as $\cos \frac{a}{2}$. One solution corresponds to the local minimum and one to the maximum. Standard algebra leads to
\begin{align}
x_{\pm} =  \frac{\alpha_0 \alpha_1 \pm\sqrt{\frac{\alpha_1^2 +4-4\alpha_0^2}{2-\alpha_0^2}} }{2(1-\alpha_0^2)}. \label{cos}
\end{align}
We have to make sure that $-1\leq x_{\pm} \leq 1$ and then we can plug both of the solutions $x_\pm$ into the Eq.~\eqref{maxEigenvalue}, simplify the results and find two candidates for the quantum value:
\begin{align}
\begin{cases}
\lambda_- &=   \frac{\alpha_0 \alpha_1 - \alpha_0^2 \sqrt{\frac{4 + \alpha_1^2 - 4\alpha_0^2}{2-\alpha_0^2}} }{1-\alpha_0^2}, \\
\lambda_+ &=   \frac{\alpha_0 \alpha_1 + \sqrt{(4 + \alpha_1^2 - 4\alpha_0^2)(2-\alpha_0^2)} }{1-\alpha_0^2}.  
\end{cases}
\end{align}
Because square roots are always non-negative $\lambda_+ \geq \lambda_-$, therefore $\lambda_+$ is a candidate for the quantum value. 
\paragraph{Summary}
If we have a Generalised Wolfe-Yelin functional with $\alpha_0 \in (-1,1)$ and $\alpha_1 \in [0,2]$, then it gives a quantum advantage when the following conditions are satisfied:
\begin{align}
\beta_Q &= \lambda_+,\\
\frac{\alpha_1}{2} &< x_+ < 1, \label{C3}\\ 
\beta_L  &< \beta_Q.  \label{C4}
\end{align}

This result is equivalent to the analytic solution of Wolfe--Yelin when $\alpha_0 \equiv x$ and $\alpha_1 \equiv -x$.

Note that $x_+$ is an explicit function of $\alpha_{0}, \alpha_{1}$, so these conditions depend only on the parameters of the functionals as desired. Numerically we have found that the first inequality given in Eq.~\eqref{C3} is satisfied when all other conditions are, but we do not have an analytical proof. A region where these conditions are satisfied is shown in Fig. \ref{regionWolf}.
\paragraph{Optimal state}
For the optimal choice of angles identifying the optimal state turns out to be easy. An explicit calculation yields
\begin{widetext}
\begin{align}
\ket{\psi} &= \cos \frac{\theta}{2} \ket{00} + \sin \frac{\theta}{2} \ket{11},\\
\cot \frac{\theta}{2} &= \frac{\alpha_1 \sqrt{2 - \alpha_0^2} + \sqrt{4 + \alpha_1^2 - 4 \alpha_0^2}(1 + \alpha_0 - \alpha_0^2)}{\sqrt{-2 \alpha_0 \alpha_1 \sqrt{(2 - \alpha_0^2)(4 + \alpha_1^2 - 4\alpha_0^2)} + \alpha_1^2(-1 - 2\alpha_0^2 + \alpha_0^4) - 4(-1 + 4 \alpha_0^2 - 4 \alpha_0^4 + \alpha_0^6)}},
\end{align}
\end{widetext}
where $\cot \frac{\theta}{2}$ is properly defined in the region specified by inequality \eqref{C3} (checked numerically).
Then, the corresponding quantum point is given by:

\begin{align}
a_0 &= a, \quad a_1 = -a, \quad b_0 = 0, \quad b_1 = \frac{\pi}{2}\\
P &= \begin{pmatrix}
\cos a \cos  \theta\\
\cos a \cos  \theta\\
\cos \theta \\
0\\
\cos a\\
\sin a \sin  \theta\\
\cos a\\
- \sin a \sin \theta
\end{pmatrix} \label{wolfePoints}
\end{align}
As shown in Appendix~\ref{Apppi2} this probability point is the unique maximiser of this Bell functional, hence, it must be an exposed point of the quantum set. We have also made a numerical observation, that one can expand the range of the parameter $a$ by discarding the condition~\eqref{C4} and obtain a larger family of extremal quantum points. However, those additional points are not exposed by any of the Generalised Wolfe--Yelin functionals, so we do not have an analytical proof of their extremality.

\subsection{Ishizaka's set of conditions for quantum points' extremality}
In a sequence of works Ishizaka developed a collection of analytic conditions for determining whether a quantum point is extremal \cite{ishizakaCryptographicQuantumBound2017, ishizakaNecessarySufficientCriterion2018, ishizakaGeometricalSelftestingPartially2020}. More specifically, this is a procedure to which we submit a quantum point and which tells us whether this point is an extremal point of the quantum set. There is no analytic proof that this procedure produces the correct result, but there is some numerical evidence to support this claim. We have tried to clarify his conditions and understand their physical meaning.

The procedure of Ishizaka consists of three conditions: the existence of a two-qubit realisation, the scaled TLM condition (STLM) and the 'greater solutions equality' condition. Let us now briefly discuss each of them.
\paragraph{Two-qubit realisation} \label{lemma2q}
The first condition demands that the specified quantum point has a two-qubit realisation. We know that all extremal quantum points have a canonical two-qubit realisation, so we can, without loss of generality, restrict our attention to this kind of quantum points. In Ref.~\cite{ishizakaNecessarySufficientCriterion2018}, Ishizaka provides a set of analytic conditions that are necessary and sufficient for a quantum point to have a canonical two-qubit realisation. For convenience, we have reformulated it as a stand-alone lemma (see Appendix~\ref{App2q} for proof).
\begin{restatable}[Existance of a canonical two-qubit realisation]{lemma}{twoqubit}
\label{Lem2Q}

Consider a real vector $P = \left( \langle A_x \rangle,  \langle B_y \rangle, \langle A_x B_y\rangle \right) \in \R^8$ satisfying $\max\{|\langle A_x \rangle|,  |\langle B_y \rangle| \} > 0$. Then, $P$ admits a two-qubit realisation of the canonical form given in Section~\ref{SecExtremal} based on a state corresponding to angle $\theta$ if and only if the following conditions are satisfied:
\begin{enumerate}
\item $z = \sin^2 \theta$ is a solution to the following equation for all pairs $(x,y)$:
\begin{align}
&z^2 -z (\langle A_x B_y\rangle^2 - \langle A_x\rangle^2 - \langle B_y \rangle^2 +1)  \nonumber\\
&+ (\langle A_x B_y\rangle - \langle A_x\rangle \langle B_y\rangle)^2 =0,\label{quadratic}
\end{align}
\item $\langle A_x \rangle ^2 \leq 1 - z$ for $x \in \{0,1\}$ and $\langle B_y \rangle ^2 \leq 1 - z$ for $y \in \{0,1\}$,
\item the following inequality holds:
\begin{align}
\prod_{x,y} \left[ \langle A_x B_y \rangle - \frac{\langle A_x \rangle\langle  B_y \rangle}{1-z} \right] \geq 0. \label{productCondition}
\end{align}
\end{enumerate}
\end{restatable}
This lemma tells us whether for a given point, there exists a canonical two-qubit realisation. Also, it can be reformulated such that one can find this realisation because as it can be seen in Appendix \ref{App2q}, the proof is constructive.  It turns out that there exist points that have more than one two-qubit realisation. This case is also discussed in Appendix \ref{App2q}.
\paragraph{Scaled Tsirelson-Landau-Masanes inequality} The second condition is a generalisation of the TLM inequality \cite{tsirelsonQuantumAnaloguesBell1987, landauEmpiricalTwopointCorrelation1988, masanesExtremalQuantumCorrelations2005}.  For a fixed quantum realisation in the CHSH scenario let us introduce the following quantities

\begin{align}
D_x^B &:= \max_{\langle \I \otimes H_B \rangle = 1} \langle A_x \otimes H_B \rangle,\\
D_y^A &:= \max_{\langle  H_A \otimes \I \rangle = 1} \langle H_A \otimes B_y \rangle.
\end{align}
where $A_x, B_y$ are observables of Alice and Bob and $H_A, H_B$ are any hermitian operators.  Let us observe that these quantities are defined only for the quantum points. It can be shown that for two-qubit canonical realisations  $D_x^B$ and $D_y^A$ can be written as follows:
\begin{align}
(D_x^B)^2 &= \langle B_y \rangle^2 + \sin^2 \theta,\\
(D_y^A)^2 &= \langle A_x \rangle^2 + \sin^2 \theta.
\end{align}
Then, Ishizaka has shown that all quantum realisations obey two following inequalities:
\begin{align}
| \tilde{C}_{00} \tilde{C}_{01} -  \tilde{C}_{10} \tilde{C}_{11}| &\leq (1 -\tilde{C}^2_{00} )^{\frac{1}{2}}(1 -\tilde{C}^2_{01} )^{\frac{1}{2}} \nonumber\\
&+ (1 -\tilde{C}^2_{10} )^{\frac{1}{2}}(1 -\tilde{C}^2_{11} )^{\frac{1}{2}} ,\label{STLM}
\end{align}
where $\tilde{C}_{xy} = \frac{\langle A_x B_y \rangle}{D_x^B}$ or $\tilde{C}_{xy} = \frac{\langle A_x B_y \rangle}{D_y^A}$. The only difference between TLM and STLM are the normalisation factors $D_x^B$ or $D_y^A$ applied to the correlators.

The second condition of Ishizaka demands that the quantum point saturates both inequalities. 
\paragraph{Greater solutions equality}Eq.~\eqref{quadratic} is a quadratic equation so it has at most two solutions. Let us denote them by $z_{xy}^\pm$ (in Ishizaka's papers they are denoted by $S_{xy}^\pm$). The last  condition demands that
\begin{align}
z = z_{00}^+ = z_{01}^+= z_{10}^+= z_{11}^+. \label{Satoshi3c}
\end{align}  
This condition was stated as a numerical observation without any physical explanation.

Now, let us gather all conditions by Ishizaka into a single conjecture.
\begin{con} \label{SatoshiCon}
Let $P$ be a quantum point with two-qubit realisation in the canonical form given in section \ref{SecExtremal}, specified by parameters $(\theta,a_0,a_1,b_0,b_1)$. Then the point $P$ is an extremal point of the quantum set if and only if it satisfies two conditions:
\begin{enumerate}
\item Point $P$ saturates the STLM criterion Eq.~\eqref{STLM} for both $D_x^B$ and $D_y^A$.
\item $z_{00}^+ = z_{01}^+= z_{10}^+= z_{11}^+$, where
\begin{align}
z_{xy}^+ &= \frac{1}{2}\big(b_{xy} + \sqrt{b_{xy}^2 - 4c_{xy}^2} \big),\\
b_{x,y} &:=  \langle A_x B_y\rangle^2 - \langle A_x\rangle^2 - \langle B_y \rangle^2 +1,\\
c_{x,y} &:=  \langle A_x B_y\rangle - \langle A_x\rangle \langle B_y\rangle.
\end{align}
\end{enumerate}
\end{con}

\subsection{New hypothesis for extremality of quantum points} \label{hypo}
By considering a large number of randomly chosen points with two-qubit realisations, we have numerically confirmed the conjecture of Ishizaka. However, we have also noticed that precisely the same results are reproduced by a distinct set of conditions.
\begin{con} \label{Ourcon}Let $P$ be a quantum point with two-qubit realisation in the canonical form given in section \ref{SecExtremal}, specified by parameters $(\theta,a_0,a_1,b_0,b_1)$. Then the point $P$ is an extremal point of the quantum set if and only if it satisfies two conditions:
\begin{enumerate}
\item The point obtained from the realisation $(\frac{\pi}{2} ,a_0, a_1, b_0, b_1)$ saturates the TLM criterion: 
\begin{align}
| C_{00} C_{01} -  C_{10} C_{11}| &=(1 -C^2_{00} )^{\frac{1}{2}}(1 -C^2_{01} )^{\frac{1}{2}} \nonumber\\
&+ (1 -C^2_{10} )^{\frac{1}{2}}(1 -C^2_{11} )^{\frac{1}{2}} ,
\end{align}
where $C_{xy} := \langle A_x B_y \rangle$.
\item $\sin \theta \geq \sin \theta^*$ where
\begin{align}
\sin \theta^* = \max\left\{  \frac{\sin a_x \sin b_y}{1-\cos a_x \cos b_y },  - \frac{\sin a_x \sin b_y}{1+\cos a_x \cos b_y }\right\} \label{Hypo2c}
\end{align}
Where a maximum is taken over all pairs $x,y$, such that $|\cos a_x \cos b_y| \neq 1$. 
\end{enumerate}
\end{con}
Our conjecture admits an elegant physical interpretation as it shows that the two conditions needed for extremality, i.e.~arrangement of observables and the entanglement of the state can be partially decoupled. The first condition determines only if the arrangmenet of observables allows extremality. Then the second condition checks whether there is enough entanglement in the state. The first condition is intuitive because we know that the TLM criterion is sufficient in cases with maximal entanglement. If for some fixed observables $\theta = \frac{\pi}{2}$ does not produce an extremal point, then smaller values of $\theta$ should not do so either.

\begin{figure}[hbtp]
\centering
\includegraphics[scale=0.5]{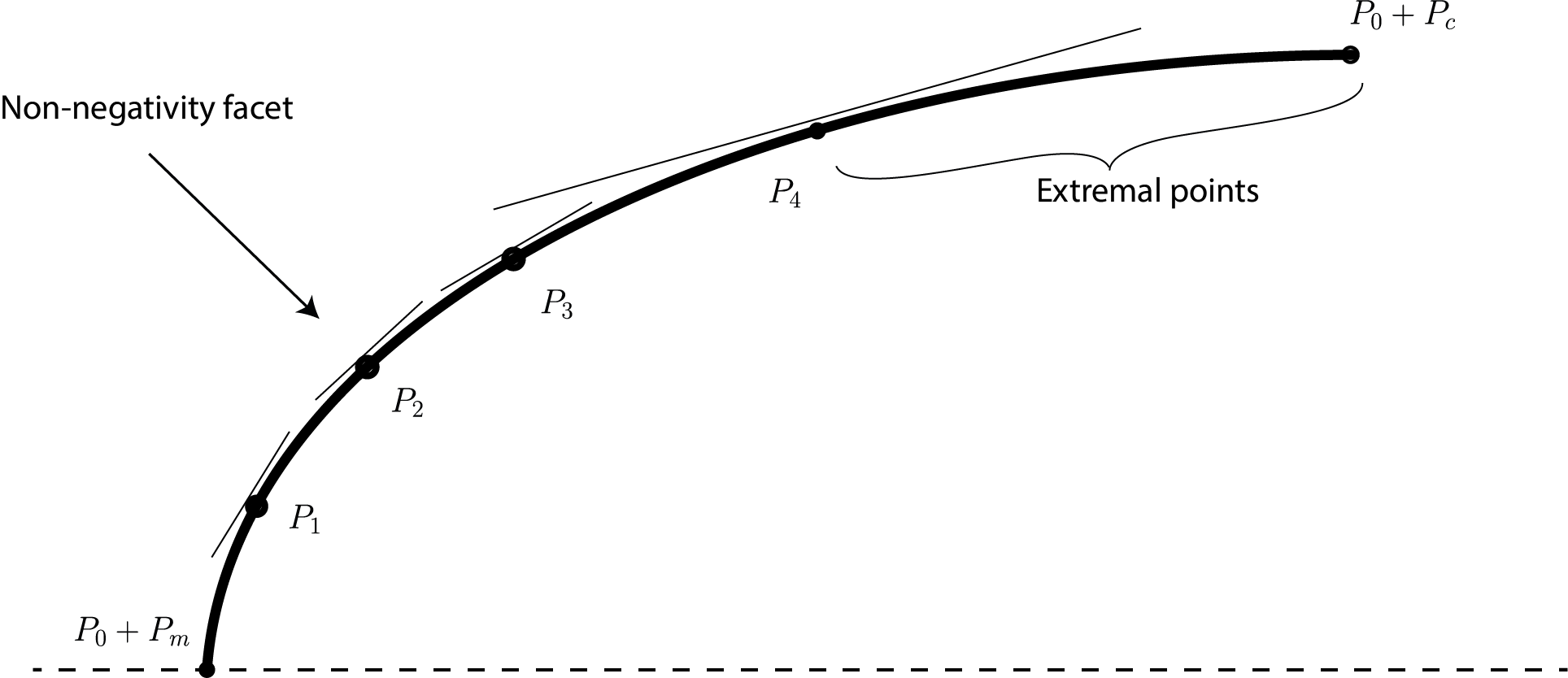}
\caption{Ilustration of conditions from Conjecture~\ref{Ourcon}. Points $P_0, P_c, P_m$ are fixed with observables' parameters. Point $P_0+P_c$ corresponds to the point with the maximally entangled state from the first condition. If this point satisfies TLM criterion, then the second condition claims that all points on the ellipse down to the first non-negativity facet (in this case $P_4$) are extremal, while points on the ellipse below $P_4$ are not extremal.} \label{elipsa}
\end{figure}

The second condition gives us the analytic value of the threshold $\theta^*$.  This value has a geometrical explanation as it can be seen in Fig.~\ref{elipsa}.  Let us write the given two-qubit point $P$ in the following way:
\begin{align}
P(\theta) = P_0 + \cos \theta P_m + \sin \theta P_c,
\end{align}
where
\begin{align}
P_0 &:= (0,0,0,0, \cos a_0 \cos b_0, \cos a_0 \cos b_1, \nonumber\\
&,\cos a_1 \cos b_0, \cos a_1 \cos b_1 ),\\
P_m &:= (\cos a_0,  \cos a_1, \cos b_0, \cos b_1,0,0,0,0 ),\\
P_c &:= (0,0,0,0, \sin a_0 \sin b_0, \sin a_0 \sin b_1 \nonumber\\
&,\sin a_1 \sin b_0,  \sin a_1 \sin b_1 ).
\end{align}
Then we can see that for fixed observables points $P_0, P_m, P_c$ are fixed and changing the parameter $\theta$ causes moving along the ellipse.  As it is shown in Fig~\ref{elipsa}, there are at most four points where this curve touches non-negativity facets (points $P_1, P_2, P_3, P_4$). The point that corresponds to the greatest value of $\theta$ ($P_4$) refers to the threshold point, which means that all points on the right-hand side are extremal and others are not.  The fact that $\theta^*$ corresponds to the point $P_4$ is shown in Appendix \ref{AppNonneg}.

In the special case, when $|\cos a_x \cos b_y| = 1$ for some pair $x, y$, we can analytically show that points with $\sin \theta < \sin \theta^*$ are not extremal. The proof of this fact is shown in Appendix~\ref{AppFace}.
\subsection{Comparison of two approaches}
In this section, we would like to compare the conditions stated by Ishizaka and ours. Firstly in both approaches, there is an assumption that a given point has a two-qubit realisation. Then the first conditions of Conjectures \ref{SatoshiCon}, \ref{Ourcon} are two different generalisations of the TLM condition. Ishizaka introduces new quantities $D_x^B, D_y^A$ and using them scales the inequality. We stay with the original TLM inequality and verify only if observables can give extremality for the maximally entangled state. It is important to mention that Ishizaka's new quantities are well-defined only for quantum points, which makes them unsuitable for the purpose of determining the boundary of the quantum set.

The second conditions Eq.~\eqref{Satoshi3c} and Eq.~\eqref{Hypo2c} look differently, but in fact, they are equivalent. It is related to the fact that the discriminant of the quadratic equation \eqref{quadratic} is equal to the product of probabilities $\Delta_{xy} = p(00|xy)p(01|xy)p(10|xy)p(11|xy)$ and it vanishes when the point lies on the non-negativity facet (see Appendix \ref{AppNonneg}).  The proof of the equivalence is provided in Appendix \ref{AppEquivalence}.

It turns out that STLM and TLM conditions are not equivalent. We verified it numerically, and for most of the cases they give the same answer, but there are points for which either STLM condition is satisfied and TLM not, or inversely. However, if one adds the last condition (which is equivalent in both approaches), then the results are precisely the same.

The advantage of our new approach is that there is clear intuition about all conditions.  For fixed observables, the criterion for extremality reduces to the elegant formula Eq.~\eqref{Hypo2c} for a threshold angle $\theta^*$. Also, our new approach can be graphically visualised as in Fig~\ref{elipsa}.

\subsection{Testing both approaches with new families of quantum extremal points}
In this section, we apply Ishizaka's and our conditions for extremality to the two new families of quantum extremal points that were derived in Section \ref{families}. Both families have two free parameters. In the case of Double-Tilted CHSH and Generalised Wolfe--Yelin families, we defined families of functionals not families of quantum points. However, every functional that uniquely exhibits quantum advantage corresponds to the quantum extremal point. These points we call respectively Double-Tilted CHSH points and Generalised Wolfe--Yelin points. We take points from a whole range of parameters of a given family and then check if they are extremal, using all of the methods: analytical conditions, numerical tools, Ishizaka's and our conditions. It turns out that we did not find any inconsistency in the results from all methods.
\subsubsection{Double-Tilted CHSH}
\begin{figure}[h]
\centering
\includegraphics[scale=0.5]{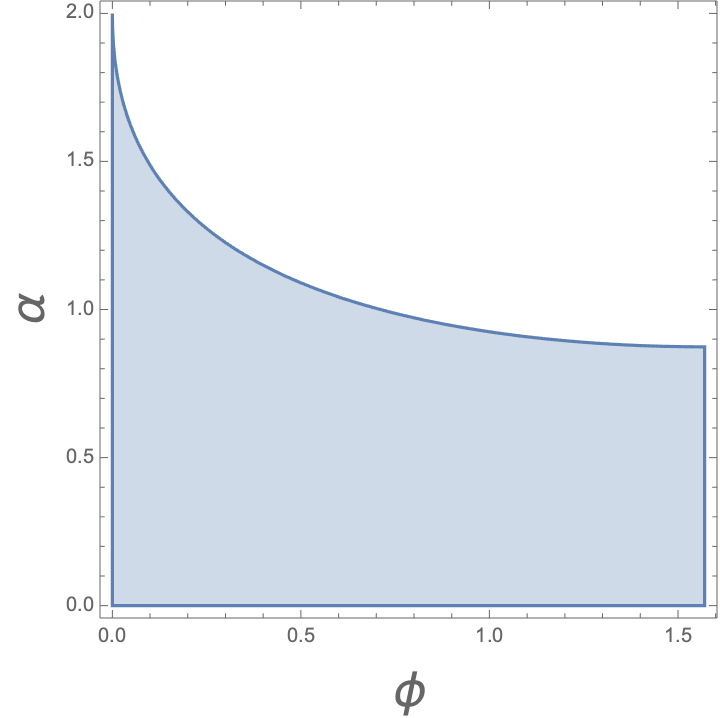}
\caption{The region of the quantum advantage of Double-Tilted CHSH functional (marked with blue colour) is exactly the same when applying analytical conditions, numerical tools, Ishizaka's and our conditions to corresponding quantum points.} \label{regionTilted}
\end{figure}

Here, we consider functionals derived in the section \ref{doubleTilted}. For points that correspond to functionals with quantum advantage, we do not have an analytical formula, but we can find it with numerical calculation. We take parameters $\phi, \alpha$ from appropriate ranges $\phi \in [0, \frac{\pi}{2}], \alpha \geq 0$. The results are shown in Fig.~\ref{regionTilted}. In this case, all conditions confirm that functionals from the blue region correspond to the quantum extremal points.
\subsubsection{Generalised Wolfe--Yelin}
\begin{figure}[h]
\centering
\includegraphics[scale=0.5]{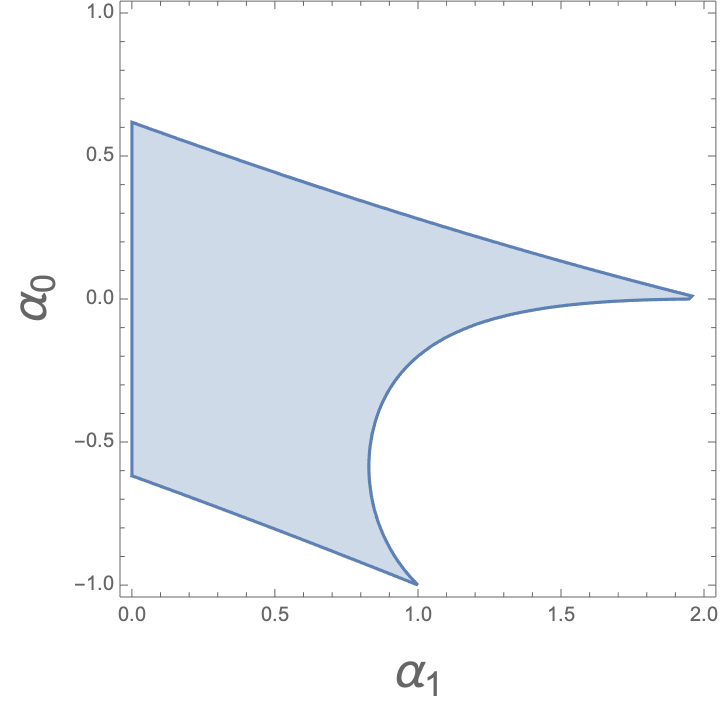}
\caption{The Generalised Wolfe--Yelin functionals with quantum advantage.} \label{regionWolf}
\end{figure}

\begin{figure}[h]
\centering
\includegraphics[scale=0.45]{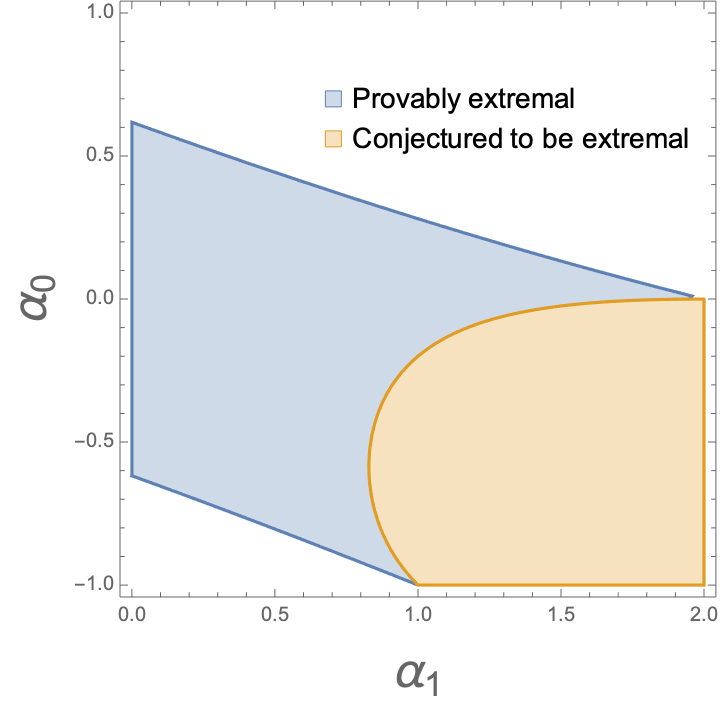}
\caption{The Generalised Wolfe--Yelin points with extended range of parameters $\alpha_0, \alpha_1$. The points that correspond to functional with quantum advantage are shown in blue. A region with quantum extremal points given by numerical tools, Ishizaka's and our conditions is shown in orange.} \label{regionsWolf}
\end{figure}

In this part, we discuss functionals from Section \ref{generalWolf}. We know the explicit formula for the representation of the corresponding family of quantum points given in Eq.~\eqref{wolfePoints}, which depends on parameters $\alpha_0, \alpha_1$. These points are defined only in the blue region in Fig.~\ref{regionWolf}, but the formula can be generalised for a wider range of $\alpha_0, \alpha_1$. The result is shown in Fig.~\ref{regionsWolf}. It turns out that all points that correspond to functionals with quantum advantage are extremal according to all methods. Furthermore, the orange region represents additional points which do not correspond to any Generalised Wolfe--Yelin functional, but with numerical tools and Conjectures \ref{SatoshiCon}, \ref{Ourcon} we found out that they are extremal.
\subsection{Extremal non-exposed points of the quantum set}
There is a specific class of extremal points that are not exposed. It means that such a point cannot be decomposed into two other quantum points, but it also does not uniquely maximise any functional. There are a few analytically known points of this class: Hardy's point \cite{hardyQuantumMechanicsLocal1992},  points of classes 3b, 2b, 2c from Ref.~\cite{chenQuantumCorrelationsNosignaling2022} and a two-parameter family of generalised Hardy's points from  Eq.~(4) in Ref.~\cite{raiSelftestingQuantumStates2022}. We checked how our Conjecture \ref{Ourcon} performs on these points and we found that they all satisfy the first condition and the second one is saturated i.e.~$\sin\theta=\sin\theta^*$. This is an intuitive and promising result because these points lie in the vicinity of exposed and boundary not extremal points, while in terms of our conjecture, they separate extremal points from not extremal ones as can be seen in Fig.~\ref{elipsa}, where they would correspond to point $P_4$. This analysis suggests that points which satisfy our first condition and saturate the second one, are extremal but not exposed. However, this is not the only manner in which we can obtain extremal not exposed points. For instance such points can be obtained from the maximally entangled state, i.e.~$\theta=\pi/2$ for example point of class 2a from \cite{chenQuantumCorrelationsNosignaling2022}. If we take such a realisation, keep the observables constant but decrease $\theta$ down to $\theta^*$, then all such points seem to be extremal and non-exposed.
\section{Conclusions}
In this work we have made two contributions to the understanding of the quantum set in the CHSH scenario. First, we have introduced an analytical description of two new two-parameter families of quantum extremal points, namely: Double-Tilted CHSH and Generalised Wolfe-Yelin. Second, we have proposed and numerically verified Conjecture~\ref{Ourcon}. This hypothesis consists of two conditions that determine whether a given probability point is an extremal point of the quantum set or not. This part was inspired by the conditions proposed by Ishizaka. We discuss his conditions and compare them to ours. We test both approaches in practice by applying them to the new families of quantum extremal points and we find that both approaches give consistent results. Although we were unable to provide a proof of our conjecture, we see the performed analysis as a strong indication. Our conjecture is related to an ellipse in a two-dimensional space (see Fig.~\ref{elipsa}), which gives us some geometric intuition, but might also constitute a promising direction for the proof. In fact, the interpretation of our conjecture is that the extremality is monotonic with the amount of the entanglement which is equivalent to the following pair of statements: (a) \emph{If a given point $P_1=(\theta_1, a_0,a_1,b_0,b_1)$ is an extremal point of the quantum set, then point $P_2=(\theta_2, a_0,a_1,b_0,b_1)$ is extremal as long as $\theta_2 \geq \theta_1$} and (b) \emph{If a given point $P_1=(\theta_1, a_0,a_1,b_0,b_1)$ is not an extremal point of the quantum set, then point $P_2=(\theta_2, a_0,a_1,b_0,b_1)$ is not extremal for any $\theta_2 \leq \theta_1$}.

Our conjecture provides an elegant separation between the arrangement of measurements and the entanglement contained in the state. Intuitively, if the measurements are sufficiently incompatible, then extremality is a monotonic function of the amount of entanglement. It would be interesting whether an analogous statement remains true in higher dimensional scenarios. In order words we should investigate what happens to nonlocality if we keep the Schmidt basis unchanged but increase the amount of entanglement. Clearly, the statement would have to be more complicated as bipartite pure states cannot be fully ordered with respect to entanglement, but one could explore the partial order based on majorisation or even a simplified approach based on comparing to the maximally entangled state. We believe this is an interesting research direction, which would enhance our understanding of the foundational aspects of Bell nonlocality.
\section{Acknowledgement}
We would like to express our gratitude to professor Yeong-Cherng for his valuable suggestion to study analytically known quantum extremal non-exposed points with our conjecture.

We acknowledge support from the National Science Centre, Poland under the SONATA project ``Fundamental aspects of the quantum set of correlations'' (grant no.~2019/35/D/ST2/02014).

\newpage
\appendix
\section{The simplest canonical parametrisation of all two-qubit quantum extremal points} \label{AppParam}
We would like to prove that Eqs.~\eqref{paramI}--\eqref{paramII}  constitute an appropriate parametrisation of all extremal quantum points. 
Here we assume that it is clear that in the CHSH scenario, every quantum extremal point has realisation with pure two-qubit state and projective observables \cite{masanesExtremalQuantumCorrelations2005}. Then, the most general parametrisation can be written as follows:
\begin{align}
\ket{\psi} &= \cos \frac{\theta}{2} \ket{00} + \sin \frac{\theta}{2} \ket{11},  \label{state}\\
A_x &= a_x^1 \I + a_x^2 \Z +  a_x^3 \X + a_x^4 \Y, \\
B_y &= b_y^1 \I + b_y^2 \Z +  b_y^3 \X + b_y^4 \Y,
\end{align}
where $a_x^i, b_y^j \in [0,2\pi)$ for all $x,y,i,j$ and $\theta \in [0,\frac{\pi}{2}]$. Hence the aim is to exclude terms with $\I$ and $\Y$. Because $A_x$ and $B_y$ are projectors, they have eigenvalues equal to $\pm 1$, hence their traces $\Tr A_x, \Tr B_y$ can take the following values: $\{ \pm 2, 0\}$. If trace of any observable takes value $\pm 2$, then the observable can be composed only from a term with $\I$ and the corresponding point would be local. To avoid that, in the parametrisation of observables there cannot be terms with $\I$ because Pauli matrices are traceless.  

Now let us focus on the terms with $\Y$. In order to show, that they have to vanish, we will follow few steps of implications.  In the section \ref{methodI} we showed, that every functional can be maximised by points with observables in the plane $\X$-$\Z$. The cost of that was the fact, that we do not know the form of the state.  Let us now argue, that this allows us to show, that we can find all extremal quantum points with parametrisation of observables Eqs.~\eqref{SecondParam1}--\eqref{SecondParam2} and then by apply unitary obtain the parametrisation Eqs.~\eqref{paramI}--\eqref{paramII}. We first have an argument that applies to exposed points and then we extend it to extremal points.
\begin{enumerate}
\item By definition of the exposed point: every exposed point of the quantum set is exposed by a certain functional.
\item For every functional one can perform the procedure described in the section  \ref{method1}. As a result, one obtains the quantum value, quantum state and observables. Then we can find the coordinates of the quantum point.
\item It turns out, that the optimal state that we obtain can be chosen to have real coefficients. It is because matrix $\bW$ from Eq.~\eqref{BellOperator} is hermitian and it has only real components. This implies, that one can choose its basis in which the eigenvectors have also only real coefficients.
\item The Schmidt bases of a real vector can be chosen to be real. This fact can be shown constructively by finding the Schmidt basis. The Schmidt basis vectors can be found by taking eigenvectors of the reduced states $\rho_B = \Tr_A \rho$, $\rho_A = \Tr_B \rho$. Because state $\rho$ has real coefficients, its reduced states also have real coefficients. Therefore we can choose their eigenvectors to be real.

\item Since every real basis on a qubit can be mapped onto the computational basis by a rotation around the $\Y$ axis, applying such a rotation on each local system will give a realisation where the state is of the canonical form, while the observables remain in the $\X$--$\Z$ plane.
\end{enumerate}

Now, we know that parametrisation Eqs.~\eqref{paramI}--\eqref{paramII} is enough to describe all quantum exposed points. There remains to show, that this parametrisation describes not only exposed points but also the extremal ones. This part of proof we devide into following steps:
\begin{enumerate}
    \item From Straszewicz's theorem \cite{basuConvexSetsMinimal2017}, we know, that toward each extremal point, converge a certain family of exposed points. Let us choose any extremal quantum point $P_e$ and the sequence of exposed quantum points $\{P_j\}$, that converge to $P_e$. 
    \item Every exposed quantum point can be described by $5$ real numbers with the two-qubit canonical realisation.
    \item From the Bolzano-Weierstrass theorem: Every sequence in a closed and bounded set $S \in \R^n$ has a convergent subsequence (which converges to a point in $S$).
    \item Hence from $\{P_j\}$ we can choose some subsequence of realisations that converges in the $P_e$.
    \item Because the map $(a_x,b_y, \theta)\rightarrow P$ is continuous, the limit of this subsequence gives us an appropriate two-qubit canonical realisation for the quantum extremal point $P_e$.
\end{enumerate}

\section{Verification of the Bell operator's eigenvalues} \label{AppEigen}
We want to find a way to verify if stated eigenvalues Eq.~\eqref{EigenMax} are appropriate eigenvalues of the Bell operator Eq.~\eqref{bellop2}.

Firstly let us prove that the non-zero eigenvalues of this operator come in pairs $\lambda, -\lambda$. In order to show that, we can observe that the Bell operator Eq.~\eqref{bellop2} has a  property that if we rotate it using unitary transformation $\Y\otimes\I$, then we obtain the negative $\bW$.
\begin{align}
(\Y\otimes\I)\bW (\Y\otimes\I) = - \bW.
\end{align}
Then we can consider the eigenequation:
\begin{align}
\bW \ket{\psi} &= \lambda \ket{\psi},
\end{align}
and define $\ket{\psi'}:= (\Y\otimes\I)\ket{\psi}$, hence
\begin{align}
\bW \ket{\psi'} = -\lambda \ket{\psi'}.
\end{align}
Therefore we can conclude that the spectrum takes the form $\{ \lambda_1, \lambda_2, -\lambda_1, -\lambda_2\}$ for some $\lambda_1, \lambda_2 \geq 0$. The argument works for non-zero eigenvalues, but if there are any zero eigenvalues, then there is even number of them. It is because from the argument above we know, that there is even number of non-zero eigenvalues and we have $4$ eigenvalues in total.

Now, let us write the matrix $\bW$ in its eigenbasis 
\begin{align}
\mathbf{W} = \sum_{i=1}^4 \lambda_i \ket{e_i}\bra{e_i}.
\end{align}
Then we can write a trace of any power of the $\mathbf{W}$
\begin{align}
\Tr(\mathbf{W}^k) = \sum_{i=1}^4 \lambda_i^k .
\end{align}
Since the non-zero eigenvalues come in pairs $\Tr(\mathbf{W}^k) = 0 $ when $k$ is an odd number.  Let us write two equations for $k=2,4$
\begin{align}
\begin{cases}
\Tr(\mathbf{W}^2) &=  2(\lambda_1^2 +\lambda_2^2) \label{verifyEigen},\\ 
\Tr(\mathbf{W}^4) &= 2(\lambda_1^4 +\lambda_2^4),
\end{cases}
\end{align}
These equations are easy to check in our specific case because it is basic algebra.  Now, our goal is to prove, that these equations have a unique solution for $\lambda_1, \lambda_2$. It is because we obtain the method for simply verifying the correctness of stated eigenvalues.

Let us formally state our problem:
Prove that for fixed $m_2,m_4$,
\begin{align}
\begin{cases} \label{unique}
m_2 &:=  2(\lambda_1^2 +\lambda_2^2), \\ 
m_4 &:= 2(\lambda_1^4 +\lambda_2^4),
\end{cases}
\end{align}
there are at most two solutions of this set of equations, over $\lambda_1, \lambda_2 \geq 0$. If there are two solutions, then they are related by swapping $\lambda_1$ and $\lambda_2$. 

Now, let us introduce the following parametrisation:
\begin{align}
\lambda_1 &= \sqrt{\frac{m_2}{2}} \cos t,\\
\lambda_2 &= \sqrt{\frac{m_2}{2}} \sin t,
\end{align}
for $t \in [0,\frac{\pi}{2}]$.  This range is sufficient, because both $\lambda_1,\lambda_2$ are non-negative numbers.  Now, we can see, that the first equation in Eqs~\eqref{unique} is satisfied for all $t$. When we plug the parametrisation into the second equation, then we obtain
\begin{align}
m_4 &= 2\left(\left(\sqrt{\frac{m_2}{2}} \cos t\right)^4 +\left(\sqrt{\frac{m_2}{2}} \sin t\right)^4\right),\\
m_4 &= 2\left(\frac{m^2_2}{4} \cos^4 t +\frac{m^2_2}{4} \sin^4 t \right),\\
\frac{2 m_4}{m_2^2} &= \cos^4 t + \sin^4 t,\\
\frac{2 m_4}{m_2^2} &=  (1- \sin^2 t)^2 + \sin^4 t,\\
\frac{2 m_4}{m_2^2} &=  1-2 \sin^2 t + 2\sin^4 t,\\
0 &=   \sin^4 t -  \sin^2 t + \frac{1 - \frac{2 m_4}{m_2^2}}{2}.
\end{align} 
This is the quadratic equation for $\sin^2 t$, hence: 
\begin{align}
\sin^2 t &= \frac{1 \pm \sqrt{4 \frac{m_4}{m^2_2} - 1}}{2}, \\
\sin t &= \pm \sqrt{\frac{1 \pm \sqrt{4 \frac{m_4}{m^2_2} - 1}}{2}}. 
\end{align}
But we know that $\sin t \geq 0$, so there are only two solutions for $\sin t$ and corresponding two for $\cos t$:
\begin{align}
\sin t &= \sqrt{\frac{1 \pm \sqrt{4 \frac{m_4}{m^2_2} - 1}}{2}}, \\
\cos t &=  \sqrt{\frac{1 \mp \sqrt{4 \frac{m_4}{m^2_2} - 1}}{2}}. 
\end{align}
Hence we can see, that there are at most two solutions for $\lambda_1, \lambda_2$:
\begin{align}
\lambda_1 &=  \frac{\sqrt{m_2 \mp \sqrt{4 m_4 - m^2_2}}}{2}, \\
\lambda_2 &= \frac{\sqrt{m_2 \pm \sqrt{4 m_4 - m^2_2}}}{2}.
\end{align}
However, the second solution is the same as the first one but with replacing $\lambda_1 \leftrightarrow \lambda_2$. 

Hence, to verify if the proposed eigenvalues are correct, one has to check two equations \eqref{verifyEigen}, which is a simple algebraic task. Now, let us show, that the highest eigenvalue $\lambda_1$ from Eq.~\eqref{EigenMax} is properly defined in the whole range of parameters $\alpha \in [0,2), \phi \in [0,\frac{\pi}{2}]$ and $a,b \in [0,\pi]$. Let us recall the formula for $\lambda_1$:
\begin{widetext}
\begin{align}
    \lambda_{1} &= \Big( 4 + \alpha^2 + \alpha^2 \cos (2a) \sin \phi  \\
    &+2\big[1 + 3 \alpha^2 - \cos(2 b) + \alpha^2 \cos b \cos \phi + 4 \alpha^2 \cos (2a) \sin \phi +\cos(4 a) (-1 + \alpha^2 + \cos(2 b) - \alpha^2 \cos b \cos \phi ) \big]^{\frac{1}{2}} \Big)^{\frac{1}{2}}. \nonumber
\end{align}
It is properly defined if terms under square roots are non-negative for all possible values of parameters. Let us consider the inner square root.
\begin{align}
    s_1 &:= 1 + 3 \alpha^2 - \cos(2 b) + \alpha^2 \cos b \cos \phi + 4 \alpha^2 \cos (2a) \sin \phi +\cos(4 a) (-1 + \alpha^2 + \cos(2 b) - \alpha^2 \cos b \cos \phi ).
\end{align}
We can reorganize $s_1$ to the following form:
\begin{align}
    s_1 & = \big(1 - \cos(4 a)\big)\big( - \alpha^2( 1 - \cos b \cos \phi)  + (1 -  \cos (2b))\big) +4 \alpha^2 \big(1 + \cos(2a) \sin \phi \big).
\end{align}
Terms $\big(1 - \cos(4 a)\big)$ and $\big(1 -  \cos (2b)\big)$ are non-negative, hence we have
\begin{align}
    s_1 &\geq -\alpha^2\big(1 - \cos(4 a)\big)\big( 1 -  \cos b \cos \phi \big) +4 \alpha^2 \big(1 + \cos(2a) \sin \phi \big).
\end{align}
Since $\cos \phi \geq 0$ for all $\phi$, we can minimize the value above by taking $b=\pi$. Also, we use a formula for the cosine of the doubled argument.
\begin{align}
    s_1 &\geq -2\alpha^2\big(1 - \cos^2(2 a)\big)\big( 1 + \cos \phi \big) +4 \alpha^2 \big(1 + \cos(2a) \sin \phi \big)\\
    &= 2\Big[ \cos^2(2a)(1+\cos\phi) +2 \sin\phi \cos(2a) +1 - \cos\phi \Big]\\
    &= 2(1+\cos\phi)\Big[ \cos^2(2a) +2 \frac{\sin\phi}{1+\cos\phi} \cos(2a) +\frac{1 - \cos\phi}{1+\cos\phi} \Big]\\
    &= 2(1+\cos\phi)\big( \cos(2a) + \frac{\sin\phi}{1+\cos\phi}\big)^2\geq 0
\end{align}
\end{widetext}

Hence, we obtain the inequality $s_1 \geq 0$, which tells us that the value under the inner square root from the maximal eigenvalue from Eq.~\eqref{EigenMax} is non-negative. Let us consider the second square root:
\begin{align}
    s2 &:= 4 +\alpha^2(1+\cos(2a)\sin\phi)+2 \sqrt{s_1}\\
    &\geq 4 +\alpha^2(1+\cos(2a)\sin\phi)\geq 4 \geq 0.
\end{align}
Hence, the value under the second square root is non-negative. Therefore, $\lambda_1$ is properly defined. Also, we checked numerically that $\lambda_2$ is well defined.
\section{Optimisation the Generalised Wolfe--Yelin Bell operator over observable parameter $b$} \label{Apppi2}
In Section \ref{generalWolf} we introduced the Generalised Wolfe--Yelin Bell operator which depends on functional parameters $\alpha_0 \in (-1,1), \alpha_1 \in [0,2]$ and two parametrisation parameters $a,b \in [0,\pi]$. In order to obtain the greatest eigenvalue of the Bell operator we optimise over $a,b$, while the functional parameters are fixed. In this section we prove that if $\beta_Q>\beta_L$, then the optimal $b$ is equal to $b=\frac{\pi}{2}$ and $2\cos(\frac{a}{2}) \geq  \alpha_1$.
\subsection{Upper bound}
Now we would like to find an upper bound for the maximal eigenvalue of $\mathbf{W}$ given in Eq.~\eqref{wolfoperator}. We show that when the functional exhibits a quantum advantage, then this bound can be achieved for $b=\frac{\pi}{2}$. Then we show, that this value is attained by a unique realisation.
First of all we can rewrite matrix  $\mathbf{W}$ in the following way:
\begin{align}
\mathbf{W} = 
\begin{pmatrix}
E & G\\
G & F
\end{pmatrix},
\end{align}
where $E,F,G$ are $2\times 2$ hermitian matrices and $E,F$ are diagonal.
\begin{align}
E &:= 
\begin{pmatrix}
\alpha_1+2\cos(\frac{a}{2})(1+\alpha_0) & 0 \\
0 & -\alpha_1+2\cos(\frac{a}{2})(-1+\alpha_0)
\end{pmatrix},\\
F &:= 
\begin{pmatrix}
\alpha_1-2\cos(\frac{a}{2})(1+\alpha_0) & 0 \\
0 & -\alpha_1+2\cos(\frac{a}{2})(1-\alpha_0)
\end{pmatrix},\\
G &:= 2\sin(\frac{a}{2})
\begin{pmatrix}
\cos b  &  \sin b \\
\sin b  & -\cos b
\end{pmatrix}. \label{G}
\end{align}
Using the same block structure we write $\ket{\psi} \in \C^4$ as
\begin{align}
\ket{\psi} := \begin{pmatrix} e_0\\e_1\end{pmatrix}, \quad e_0, e_1 \in \C^2, \quad \vert e_0 \vert^2 + \vert e_1 \vert^2 = 1,
\end{align}
and use it in order to bound the maximal eigenvalue of  $\mathbf{W}$ as follows:
\begin{align}
&\bra{\psi}\mathbf{W}\ket{\psi} = (e_0^\dagger \quad e_1^\dagger)\begin{pmatrix}E & G\\G & F\end{pmatrix}\begin{pmatrix} e_0\\e_1\end{pmatrix}\\ 
&= e_0^\dagger E e_0 + e_1^\dagger F e_1 + 2 \operatorname{Re}(e_0^\dagger G e_1)  \label{estimation1}\\
&\leq e_0^\dagger e_0 \lambda_{\max}(E)  + e_1^\dagger e_1 \lambda_{\max}(F) + 2 \vert e_0^\dagger G e_1\vert  \label{estimation2}\\
&\leq \vert e_0\vert^2 \lambda_{\max}(E)  + \vert e_1 \vert^2 \lambda_{\max}(F) + 2 \vert e_0\vert \vert e_1 \vert  \Vert G \Vert _{\infty}=:\Lambda ,\label{estimation3}
\end{align}
where $\Vert G \Vert _{\infty}$ denotes the Schatten infinity norm. To obtain the maximal value of $\Lambda$ we optimise its value over vectors $ e_0, e_1$ such that $\vert e_0\vert^2 + \vert e_1\vert^2 = 1$. We can see that this problem is equivalent to finding the maximal eigenvalue of a $2 \times 2$ matrix $\mathbf{M}$ defined as follows:
\begin{align}
\bM := 
\begin{pmatrix}
\lambda_{\max}(E) & \Vert G \Vert _{\infty}\\
\Vert G \Vert _{\infty} & \lambda_{\max}(F)
\end{pmatrix}.
\end{align}
Because for hermitian operators a Schatten infinity norm is defined as the greatest absolute value of its eigenvalues it is easy to check that $ \Vert G \Vert _{\infty} = 2\sin(\frac{a}{2})$. $\lambda_{\max}(E)$ and $\lambda_{\max}(F)$ are also easy to calculate because $E, F$ are diagonal matrices.
\begin{align}
\lambda_{\max}(E) &= \alpha_1+2\cos(\frac{a}{2})(1+\alpha_0), \\
\lambda_{\max}(F) &= \max\{ -\alpha_1+2\cos(\frac{a}{2})(1-\alpha_0), \nonumber\\
&\alpha_1-2\cos(\frac{a}{2})(1+\alpha_0)\}. \label{Feigen}
\end{align}
Without any other assumptions, we cannot tell which of the two expressions given in Eq.~\eqref{Feigen} is greater. Hence we have two cases to consider:
\begin{align}
-\alpha_1+2\cos(\frac{a}{2})(1-\alpha_0)  \geq \alpha_1-2\cos(\frac{a}{2})(1+\alpha_0),  \\
-\alpha_1+2\cos(\frac{a}{2})(1-\alpha_0)  < \alpha_1-2\cos(\frac{a}{2})(1+\alpha_0). 
\end{align}
We can reduce these inequalities so they do not depend on $\alpha_0$ parameter.
\begin{align}
2\cos(\frac{a}{2}) \geq \alpha_1 \label{conditionI},\\
2\cos(\frac{a}{2}) <  \alpha_1 \label{conditionII},
\end{align}
and they correspond to the following matrices $\bM$:
\begin{align}
\bM_1 =\begin{pmatrix}
\alpha_1+2\cos(\frac{a}{2})(1+\alpha_0) & 2\sin(\frac{a}{2})\\
2\sin(\frac{a}{2}) & -\alpha_1+2\cos(\frac{a}{2})(1-\alpha_0)
\end{pmatrix},\label{MmatricesI}\\
\bM_2 =\begin{pmatrix}
\alpha_1+2\cos(\frac{a}{2})(1+\alpha_0) & 2\sin(\frac{a}{2})\\
2\sin(\frac{a}{2}) & \alpha_1-2\cos(\frac{a}{2})(1+\alpha_0)
\end{pmatrix}. \label{MmatricesII}
\end{align}
We have derived an upper bound on $\lambda_{\max}(\bW) = \max \{\lambda_{\max}(\bM_1), \lambda_{\max}(\bM_2)\}$ which only depends on $\alpha_0, \alpha_1$ and $a$. Let us show that this upper bound is tight and saturated when either $b=0$ or $b=\frac{\pi}{2}$. Points corresponding to $b=0$ are local and lead to the local value, but for $b=\frac{\pi}{2}$, they can be non-local, and for them, we can observe an advantage of a quantum value.

Let us write down the matrix $\mathbf{W}$ in the case when $b=\frac{\pi}{2}$: 
\begin{widetext}
\begin{align}
&\bW(b=\frac{\pi}{2})=\label{newWI}\\
&\begin{pmatrix}
\alpha_1+2\cos(\frac{a}{2})(1+\alpha_0) & 0 & 0 &  2 \sin(\frac{a}{2})\\
0 & -\alpha_1+2\cos(\frac{a}{2})(-1+\alpha_0) &  2 \sin(\frac{a}{2}) & 0\\
0 &  2 \sin(\frac{a}{2}) & \alpha_1-2\cos(\frac{a}{2})(1+\alpha_0) & 0\\
2 \sin(\frac{a}{2}) & 0 & 0 & -\alpha_1+2\cos(\frac{a}{2})(1-\alpha_0)\\
\end{pmatrix}.\nonumber
\end{align}
\end{widetext}
As we can see, matrix $\bW(b=\frac{\pi}{2})$ is block-diagonal with two blocks: matrix $\bM_1$  and some other matrix. Analogously, $\bW(b=0)$ splits up into the matrix $\bM_2$ and some other matrix. It means that when the condition given in Eq.~\eqref{conditionI} holds, then we know that the largest eigenvalue of $\bW$ will be obtained when $b= \frac{\pi}{2}$. We have not shown that this is the only possibility yet, but we know that for any $b$ the largest eigenvalue cannot be larger than this one with $b= \frac{\pi}{2}$. Analogously when condition from Eq.~\eqref{conditionII} holds, then we get the maximal eigenvalue of  $\mathbf{W}$ for $b=0$, but this is the local point, so we do not consider this case further.
Hence we can say that the maximal eigenvalue can be achieved when both these conditions obey:
\begin{align}
\begin{cases}
b = \frac{\pi}{2},\\
2\cos(\frac{a}{2}) \geq  \alpha_1.
\end{cases}
\end{align}
We can simplify $\bW$ to the form from Eq.~\eqref{newWI}. Then it can be separated into two matrices, and then its largest eigenvalue can be written as:
\begin{align}
\lambda_{\bW}(a) &= 2 \cos (\frac{a}{2}) + \Big( 2 + \alpha_1^2 + 2 \alpha_0^2 \nonumber\\
&+ 4 \alpha_0\alpha_1 \cos (\frac{a}{2}) - 2 \cos a(1 -  \alpha_0^2)\Big)^{\frac{1}{2}}. \label{maxEigenvalue2}
\end{align}
Hence if $\beta_Q > \beta_L$, then $\beta_Q = \max_a \lambda_\bW(a)$.
\subsection{Uniqueness}
In the case of functionals which have a quantum-classical gap, let us observe that $b = \frac{\pi}{2}$ is the only possible choice if we want to obtain extremal points. It follows from the fact that inequalities \eqref{estimation1}--\eqref{estimation3} have to be saturated. If eigenvalues of matrix $F$ given in Eq.~\eqref{Feigen} are not equal, then from the first two expressions in the inequality \eqref{estimation1}--\eqref{estimation2} we see that vectors $e_0, e_1$ have to have one of the coordinates equal to $0$. Then from the third expression in the inequality \eqref{estimation2}--\eqref{estimation3}, we can see that matrix $G$ has to take the form of the Pauli matrix $\X$. Therefore using the form of $G$ Eq.~\eqref{G} we see, that $b = \frac{\pi}{2}$ is the only possible choice, to saturate all inequalities. In case when eigenvalues of matrix $F$ are equal, we can see that matrices $M_1, M_2$ given in Eqs.~\eqref{MmatricesI},~\eqref{MmatricesII} are equivalent. It means that the derived in the previous section upper bound is saturated for both $b=0, b=\frac{\pi}{2}$ and this implies that the given functional does not exhibit a quantum advantage.
\section{Analytic conditions for existence of a two-qubit canonical realisation} \label{App2q}
In Section \ref{lemma2q} we introduced the lemma that tells us if a given point has a two-qubit canonical realisation. Now, we would like to prove it. Let us recall \cref{Lem2Q}:

\twoqubit*

In the proof, one direction is very simple to show and the second one is the real task. If the point has a two-qubit realisation in the canonical form, then it has parametrisation Eqs.~\eqref{real1}--\eqref{real2} and by simple substituting marginals and correlators we can show, that all three conditions from the lemma are satisfied. Also we should notice that $\theta \neq \frac{\pi}{2}$ because of the condition $\max\{|\langle A_x \rangle|,  |\langle B_y \rangle| \} > 0$, that tells us, that at least one of the marginals is non-zero. Now we will focus on the second part of the proof, which demands to show, that if the conditions are satisfied then the given point has two-qubit canonical realisation.

In order to prove that the given point has two-qubit realisation we will show, that for a given point (eight numbers $(\langle A_x\rangle, \langle B_y\rangle, \langle A_xB_y\rangle)$)  there exist parameters $\theta, a_x, b_y$ such that, the equations \eqref{real1}--\eqref{real2} hold.  Now, we can observe that we can compute $\sin^2 \theta$ in terms of marginals and correlators. Let us perform simple transformation of the equations \eqref{real1}--\eqref{real2} and take square of the last one.
\begin{align}
\cos a_x  &= \frac{\langle A_x\rangle}{\cos \theta}, \label{cos1}\\
\cos b_y  &= \frac{\langle B_y\rangle}{\cos \theta},\label{cos2}\\
\sin^2 \theta \sin^2 a_x \sin^2 b_y &= (\langle A_x B_y \rangle - \cos a_x \cos b_y)^2.\label{squared}
\end{align}
Now we can eliminate $\cos a_x,  \sin a_x, \cos b_y,  \sin b_y$ and obtain
\begin{widetext}
\begin{align}
(1 - \cos^2 \theta)(1 -  \frac{\langle A_x\rangle^2}{\cos^2 \theta})(1 -  \frac{\langle B_y\rangle^2}{\cos^2 \theta}) = \langle A_x B_y \rangle^2 - 2\langle A_x B_y \rangle \frac{\langle A_x\rangle}{\cos \theta}\frac{\langle B_y\rangle}{\cos \theta} + \frac{\langle A_x\rangle^2}{\cos^2 \theta}\frac{\langle B_y\rangle^2}{\cos^2 \theta}, \label{quadraticEqu}
\end{align}
hence
\begin{align}
0 &= \cos^2 \theta (1 - \cos^2 \theta)(1 -  \frac{\langle A_x\rangle^2}{\cos^2 \theta} -  \frac{\langle B_y\rangle^2}{\cos^2 \theta} + \frac{\langle A_x\rangle^2}{\cos^2 \theta} \frac{\langle B_y\rangle^2}{\cos^2 \theta}) \nonumber\\
&- \cos^2 \theta \langle A_x B_y \rangle^2 + 2\langle A_x B_y \rangle \langle A_x\rangle \langle B_y\rangle - \frac{\langle A_x\rangle^2 \langle B_y\rangle^2}{\cos^2 \theta}\\
&= \cos^2 \theta  - \langle A_x\rangle^2 - \langle B_y\rangle^2 + \frac{\langle A_x\rangle^2 \langle B_y\rangle^2}{\cos^2 \theta} - \cos^4 \theta + \cos^2 \theta( \langle A_x\rangle^2 + \langle B_y\rangle^2) - \langle A_x\rangle^2 \langle B_y\rangle^2 \nonumber\\
&- \cos^2 \theta \langle A_x B_y \rangle^2 + 2\langle A_x B_y \rangle \langle A_x\rangle \langle B_y\rangle - \frac{\langle A_x\rangle^2 \langle B_y\rangle^2}{\cos^2 \theta}.
\end{align}
Now, we want to change cosines for sines.  
\begin{align}
0 = \sin^4\theta -\sin^2\theta (\langle A_x B_y\rangle^2 - \langle A_x\rangle^2 - \langle B_y \rangle^2 +1) + (\langle A_x B_y\rangle - \langle A_x\rangle \langle B_y\rangle)^2. \label{quadratic3}
\end{align}
\end{widetext}
Hence, for each $x, y \in {0, 1}$ we obtain a quadratic equation for $z= \sin^2 \theta$.
\begin{align}
0 &= z^2 -b_{x y} z + c^2_{x y},\\
b_{x y} &:=  \langle A_x B_y\rangle^2 - \langle A_x\rangle^2 - \langle B_y \rangle^2 +1,\\
c_{x y} &:=  \langle A_x B_y\rangle - \langle A_x\rangle \langle B_y\rangle.
\end{align}
Then each equation has at most two solutions which we denote $z_{x y}^\pm$:
\begin{align}
z^{\pm}_{x y} = \frac{1}{2}\left( b_{x y} \pm \sqrt{b_{x y}^2 - 4 c^2_{x y}} \right). \label{quadraticSol}
\end{align}
We can observe that satisfying Eq.~\eqref{quadratic3} for all $x,y$ is precisely the first condition in the Lemma. Hence we can see, that this condition provides an appropriate candidate for the parameter $\theta$. Now we have to check what we can conclude knowing, that $z_{x,y} = \sin^2(\theta)$ is the solution for all equations \eqref{quadratic3}. We can observe that this is equivalent to satisfying Eq.~\eqref{quadraticEqu}. Then thanks to the second condition we can determine $\cos a_x, \cos b_y$ like in Eqs.~\eqref{cos1}--\eqref{cos2} and obtain Eq.~\eqref{squared}.  Here we come across the first issue, because Eq.~\eqref{squared} is not equivalent to Eq.~\eqref{real2}. From Eq.~\eqref{squared} we conclude that for each pair $x, y$ either
\begin{align}
\langle A_x B_y \rangle &= \cos a_x \cos b_y +  \sin \theta \sin a_x \sin b_y,
\end{align}
or
\begin{align}
\langle A_x B_y \rangle &= \cos a_x \cos b_y -  \sin \theta \sin a_x \sin b_y,
\end{align}
holds. We can easily check that if for all $x,y$ there is odd number of equations with minuses in the second term at all, then we obtain wrong parametrisation. Otherwise the parametrisation is correct but it can be defined differently equivalently up to local transformations of Alice and Bob's observables. However, here we can make use of the third condition. If we substitute into it the parametrisation with odd number of minuses, then we obtain the following inequality:
\begin{align}
- \sin^2 a_0 \sin^2 a_1 \sin^2 b_0 \sin^2 b_1 \geq 0,
\end{align}
which can be true only if at least one of the sines vanishes. However, in such cases the number of minus signs in the parametrisation is not relevant, because at least two terms with this sign vanish, and then the parametrisation is equivalent to the Eqs.~\eqref{real1}--\eqref{real2}.

We have already found $\sin^2 \theta$ and $\cos a_x, \cos b_y$. Because $\theta \in [0,\frac{\pi}{2})$,  we can find unambiguously the value of $\theta$. Then the only remaining part of the proof is to verify if one can find $\sin a_x$ and $\sin b_y$.   We can write that:
\begin{align}
\sin a_x &= \pm \sqrt{1- \cos a_x^2},\\
\sin b_y &= \pm \sqrt{1- \cos b_y^2}.
\end{align}
The last thing to do, is to find the signs of the $\sin a_x$ and $\sin b_y$. We can observe that the signs cannot be taken arbitrarily because the Eqs.~\eqref{real2} have to hold.  So let us move all known parts from this equations onto the right hand side
\begin{align}
\sin a_x \sin b_y = \frac{\langle A_x B_y \rangle - \cos a_x \cos b_y}{\sin \theta}. \label{signs}
\end{align}
Now, we have four equations for products of $\sin a_x, \sin b_y$ and the task is to check if we can choose the signs of all sines so that Eqs.~\eqref{signs} were satisfied. Let us denote right hand side of the equation as $s_{xy}$. Then we have
\begin{align}
\sin a_0 \sin b_0 &= s_{00},\\
\sin a_0 \sin b_1 &= s_{01},\\
\sin a_1 \sin b_0 &= s_{10},\\
\sin a_1 \sin b_1 &= s_{11}.
\end{align}
Now we can observe, that we can construct the algorithm  for choosing signs of the sines. Firstly we can choose that $\sin a_0 \geq 0$. This implies the signs of $\sin b_0 = \frac{s_{00}}{\sin a_0}$ and $\sin b_1 = \frac{s_{01}}{\sin a_0}$. Now, we are left with two equations for sign of the $\sin a_1$ and both of them have to obey.  So
\begin{align}
\begin{cases}
\sin a_1 &= \frac{s_{10}}{\sin b_0} = \frac{s_{10}}{s_{00}} \sin a_0,\\
\sin a_1 &= \frac{s_{11}}{\sin b_1} = \frac{s_{11}}{s_{01}} \sin a_0.\\
\end{cases}
\end{align}
Hence, we see that $\frac{s_{10}}{s_{00}} = \frac{s_{11}}{s_{01}}$. Because of the fact that this equation is satisfied up to the sign, we can see that this is equivalent to the condition that there has to be even number of negative $s_{xy}$. We can write this condition as follows $s_{00}s_{01}s_{10}s_{11} \geq 0$. Now, we want to write this condition using the marginals and correlators.
\begin{align}
s_{00}s_{01}s_{10}s_{11} &\geq 0, \\
\prod \left( \frac{\langle A_x B_y \rangle - \cos a_x \cos b_y}{\sin \theta} \right) &\geq 0, \\
\end{align}
since $\sin \theta \geq 0$, this is equivalent to
\begin{align}
\prod \left( \langle A_x B_y \rangle - \cos a_x \cos b_y \right) &\geq 0, \\
\prod \left( \langle A_x B_y \rangle - \frac{\langle A_x\rangle}{\cos \theta} \frac{\langle B_y\rangle}{\cos \theta} \right) &\geq 0, \\
\prod \left( (1 - z) \langle A_x B_y \rangle - \langle A_x \rangle \langle B_y\rangle \right) &\geq 0,
\end{align}
which is again the third condition,  therefore we can always find appropriate signs of the sines. Hence, we showed that if conditions from Lemma are satisfied, then the given point has two-qubit realisation. The proof is constructive, so we can find parameters $\theta, a_x, b_y$, that describe this realisation.
\subsection{Uniqueness}
It turns out, that it is possible, that certain points have two different two-qubit canonical realisation. This may happen, when in Eq.~\eqref{quadraticSol} $z_1=z_{xy}^+$, $z_2=z_{xy}^-$ and $z_1 \neq z_2$ for all $x,y\in \{0,1\}$. In such case, we have two appropriate and distinct values of $\sin \theta$. It is difficult to find all such points, because there is a lot of cases to consider. Let us show an example of one quantum point, that have two not equivalent two-qubit canonical realisations: 
\begin{align}
P &= \big( \frac{1}{4}, \frac{1}{2}, \frac{1}{2},\frac{1}{2},\frac{11}{16}, \frac{11}{16}, \frac{13}{16}, \frac{13}{16}\big).
\end{align}

This point gives $z_{xy}^+= 2^{-9}(297 + 9\sqrt{65}) \approx 0.72 $, $z_{xy}^-=2^{-9}(297 - 9\sqrt{65})\approx 0.44$ for all $x,y\in \{0,1\}$. Both realisations obey all conditions from Lemma \ref{Lem2Q}. However, this point is local. Using numerical calculations we found only local quantum points with two different two-qubit canonical realisations. Note, that there is also a lot of points which have $z_1=z_{xy}^+$, $z_2=z_{xy}^-$ and $z_1 \neq z_2$ for all $x,y\in \{0,1\}$, but for realisation with $\sin^2 \theta = z_2$ the condition from Eq.~\eqref{productCondition} is not satisfied, so they have only one realisation.
\section{Relation between non-negativity facets and conditions for extremality} \label{AppNonneg}
Non-negativity facet is a facet of points where one of the probabilities $p(ab|xy)$ equals 0. When a point lies on such a facet, then it is a boundary point. It turns out, that these facets play a crucial role in the second condition from Conjecture \ref{Ourcon}. In this appendix we show the connection of non-negativity facets with the mentioned condition and with the discriminant of the quadratic equation \eqref{quadratic}.
\subsection{Vanishing of the probability} \label{vanishingOfProb}
Let us find the condition for vanishing probability $p(ab|xy)$ when observables of a given point with two-qubit canonical realisation are fixed:
\begin{align}
0 &= p(ab|xy),\label{prob0}\\
0 &= (-1)^{a+b}\langle A_x B_y\rangle +(-1)^a \langle A_x\rangle + (-1)^b\langle B_y \rangle + 1,\\
0 &= (-1)^{a+b} \sin a_x \sin b_y \sin \theta + (-1)^{a+b}\cos a_x \cos b_y \nonumber\\
&+ \big((-1)^a \cos a_x + (-1)^b \cos b_y \big)\cos \theta + 1 , \label{vanishProb}
\end{align}
we can define parameters $A,B,C$
\begin{align}
A &:= (-1)^{a+b} \sin a_x \sin b_y,\\
B &:=  (-1)^a \cos a_x + (-1)^b \cos b_y, \\
C &:= 1 + (-1)^{a+b}\cos a_x \cos b_y.
\end{align}
Then Eq.~\eqref{vanishProb} reads
\begin{align}
A \sin \theta + B\cos \theta  &= -C.
\end{align}
Now we want to find $\theta \in [0,\frac{\pi}{2}]$ that fulfils above equation for given parameters $a_0,a_1,b_0,b_1$. Let us observe that
\begin{align}
&\sqrt{A^2 + B^2} = \big( \sin^2 a_x \sin^2 b_y + \cos^2 a_x +\cos^2 b_y \nonumber\\ 
&+ (-1)^{a+b}2 \cos a_x \cos b_y\big)^{\frac{1}{2}} \\
&=\big(1 + \cos^2 a_x \cos^2 b_y + (-1)^{a+b}2 \cos a_x \cos b_y\big)^{\frac{1}{2}}   \nonumber\\
&= 1 + (-1)^{a+b} \cos a_x \cos b_y = C.
\end{align}
We can use the trigonometric identity that
\begin{align}
A \sin \theta + B\cos \theta = \sqrt{A^2 + B^2} \sin(\theta + \phi),
\end{align}
where
\begin{align}
\sin \phi &= \frac{B}{\sqrt{A^2 + B^2} },\\
\cos \phi &= \frac{A}{\sqrt{A^2 + B^2} }.\\
\end{align}
Hence we get
\begin{align}
C\sin(\theta +\phi) &= -C,\\
\sin(\theta +\phi) &= -1,\\
\theta +\phi &= - \frac{\pi}{2} + 2 k\pi,\\
\sin \theta &= -\sin(\phi + \frac{\pi}{2}) = -\cos \phi = \\
&=- \frac{(-1)^{a+b} \sin a_x \sin b_y}{1 + (-1)^{a+b} \cos a_x \cos b_y},\label{sinTheta}\\
\cos \theta &= \cos(\phi + \frac{\pi}{2}) = -\sin \phi = \\
&=- \frac{(-1)^{a} \cos a_x  + (-1)^b\cos b_y}{1 + (-1)^{a+b} \cos a_x \cos b_y}.
\end{align}
There is the particular case when $C=0$ then $a_x, b_y \in \{ 0,\pi \}$. In this case, also $A=0$, and for two combinations of pair $a,b$ $B=0$. This implies, that two probabilities vanish independently of the value of $\theta$. Also let us observe that $\sin \theta \geq 0$ and $\cos \theta \geq 0$ which implies that $A,B \leq 0$. In other cases, there is no solution to the Eq.~\eqref{prob0}. 

We can note that for fixed $x, y$ there can be only one pair $a,b$ for which probability $p(ab|xy)$ vanishes.  This comes from the fact, that $\sin \theta \geq 0, \cos \theta \geq 0$ and this is possible only for one choice of $a,b$.

If we take maximum of $\sin \theta$ from Eq.~\eqref{sinTheta} over $a,b,a_x,b_y$, then we obtain exactly the formula for $\sin \theta^*$ from Eq.~\eqref{Hypo2c}.

\subsection{Discriminant in the quadratic equation}
If $\Delta_{xy}$ is the discriminant of the quadratic equation \eqref{quadratic}, then we can observe that its vanishing is related to vanishing one of the probability $p(ab|xy)$.
\begin{widetext}
\begin{align}
\Delta_{xy} &= \big(\langle A_x B_y\rangle^2 - \langle A_x\rangle^2 - \langle B_y \rangle^2 +1 \big)^2 - 4\big( \langle A_x B_y\rangle - \langle A_x\rangle \langle B_y\rangle \big)^2 \\
&=\big(\langle A_x B_y\rangle^2 - \langle A_x\rangle^2 - \langle B_y \rangle^2 +1 - 2 \langle A_x B_y\rangle +2 \langle A_x\rangle \langle B_y\rangle \big)  \\
&\cdot\big( \langle A_x B_y\rangle^2 - \langle A_x\rangle^2 - \langle B_y \rangle^2 +1 +2\langle A_x B_y\rangle -2 \langle A_x\rangle \langle B_y\rangle \big) \\
&=\Big(\big(\langle A_x B_y\rangle -1\big)^2 - \big(\langle A_x\rangle - \langle B_y \rangle\big)^2 \Big) \cdot\Big(\big(\langle A_x B_y\rangle +1\big)^2 - \big(\langle A_x\rangle + \langle B_y \rangle\big)^2 \Big)\\
&= \big(\langle A_x B_y\rangle -1 - \langle A_x\rangle + \langle B_y \rangle\big)  \big(\langle A_x B_y\rangle -1+\langle A_x\rangle - \langle B_y \rangle \big) \\
&\cdot \big(\langle A_x B_y\rangle +1 - \langle A_x\rangle - \langle B_y \rangle\big) \big(\langle A_x B_y\rangle +1+\langle A_x\rangle + \langle B_y \rangle\big)  \\
&= p(00|xy)p(01|xy)p(10|xy)p(11|xy).
\end{align}
\end{widetext}
Hence when some of the probability vanishes then $\Delta_{xy}$ vanishes as well. Also, we can see that if the given point is an appropriate point constructed from probabilities, then $\Delta_{xy} \geq 0$.
\section{Equivalence of the second conditions in Conjectures \ref{SatoshiCon},\ref{Ourcon}}\label{AppEquivalence}
In this section we prove, that the second condition from Conjecture \ref{SatoshiCon} by Ishizaka is equivalent to the second condition from our Conjecture \ref{Ourcon}. We have to show, that if and only if $\theta \geq \theta^*$ where $\theta^*$ is the threshold angle from Eq.~\eqref{Hypo2c}, then  Eq.~\eqref{Satoshi3c} holds.

We know, that $z_{xy} = \sin^2 \theta$. Therefore from Vieta's formulas second solution of the quadratic equation \eqref{quadratic} is $z^*_{xy} = b_{xy} - z_{xy}$. We can see, that $z_{xy}^+ \geq z_{xy}^-$ for all $x,y$. Hence the Ishizaka's condition is equivalent to that $z_{xy} \geq z_{xy}^*$ for all $x,y$.
\begin{align}
z_{xy} &\geq z_{xy}^*,\\
z_{xy} &\geq b_{xy} - z_{xy},\\
2 z_{xy} &\geq b_{xy}, \\
2 \sin^2 \theta &\geq \langle A_x B_y\rangle^2 - \langle A_x\rangle^2 - \langle B_y \rangle^2 +1.
\end{align}
If we plug in a two-qubit parametrisation then we obtain the following inequality:
\begin{align}
2 \sin^2 \theta \geq& (\cos a_x \cos b_y + \sin \theta \sin a_x \sin b_y)^2 \nonumber\\
&- \cos^2 a_x \cos^2 \theta - \cos^2 b_y \cos^2 \theta +1.
\end{align}
After a few transformations, we obtain a quadratic inequality for $\sin \theta$
\begin{align}
0 &\leq \sin^2 \theta (1 - \cos^2 a_x \cos^2 b_y) \nonumber\\
& - 2 \sin \theta \sin a_x \sin b_y \cos a_x \cos b_y - \sin a_x^2 \sin^2 b_y.
\end{align}
Because $\sin \theta \geq 0$, this inequality is satisfied for such $\sin \theta$ that 
\begin{align}
\sin \theta \geq \max \big\{- \frac{\sin a_x \sin b_y}{\cos a_x \cos b_y - 1}, - \frac{\sin a_x \sin b_y}{\cos a_x \cos b_y + 1} \big\}.
\end{align}

Hence we can see, that for each $x,y$ there is one threshold value of $\theta$ that for smaller one there is $z_{xy}^-= \sin^2 \theta$ but for greater $z_{xy}^+= \sin^2 \theta$. We can also see in Appendix~\ref{AppNonneg} that these values of $\theta$ corresponds to the points that lie on some non-negativity facet. Therefore if we take 
\begin{align}
\sin \theta^* =\max_{x,y}  \big\{- \frac{\sin a_x \sin b_y}{\cos a_x \cos b_y - 1}, - \frac{\sin a_x \sin b_y}{\cos a_x \cos b_y + 1} \big\},
\end{align}
then we obtain points for which $z_{00}^+ =z_{01}^+ =z_{10}^+ =z_{11}^+$ and this is precisely the formula in our second condition (Eq.~\eqref{Hypo2c}).
\section{The non-extremality of points in the particular case, when a whole ellipse lies on the non-negativity facet} \label{AppFace}
In the second condition in the hypothesis from the section \ref{hypo} there is a special case, when $|\cos a_x \cos b_y| = 1$ for some pair $x, y$. From now on, let us suppose, that $a_0 = b_0 = 0$, which implies $\cos a_0 \cos b_0 = 1$. All other cases can be considered analogously. We showed in Section~\ref{vanishingOfProb}, that in this particular case, for fixed observables $a_0, a_1, b_0, b_1$ independently of $\theta$ the point lies on the two non-negativity facets $p(00|00)= p(11|00)=0$ or $p(01|00) = p(10|00)=0$. In this section we would like to prove, that all points with these observables and $\sin \theta < \sin \theta^*$, where $\sin \theta^*$ is given in Eq.~\eqref{Hypo2c}, are not extremal.

Let us firstly argue, that we can consider only observables parameters $a_x, b_y$ from the range $[0,\pi]$. It comes from the fact, that we can always consider transformations of Alice's and Bob's observables operators $A_x \rightarrow - A_x$, $B_y \rightarrow - B_y$, that do not change the physics of the problem. These transformations are equivalent to shift of the observables parameters $a_x \rightarrow a_x \pm \pi, b_y\rightarrow b_y \pm \pi$. Hence we can consider only $a_1, b_1$ from the range $[0,\pi]$.

Now, it is easy to see, that the formula for the threshold $\theta^*$ is given by 
\begin{align}
    \sin \theta^* &= \frac{\sin a_1 \sin b_1}{1 - \cos a_1 \cos b_1},\\
    \cos \theta^* &= \frac{|\cos a_1 - \cos b_1|}{1 - \cos a_1 \cos b_1}.
\end{align}
Then, let us recall that points for fixed $a_0, a_1, b_0, b_1$ lie on the ellipse that in this case can be described as follows:
\begin{align}
    P &= P_0 + \cos \theta P_m +\sin \theta P_c,\\
    P_0 &= (0,0,0,0, 1, \cos b_1, \cos a_1, \cos a_1 \cos b_1),\\
    P_m &= (1, \cos a_1, 1, \cos b_1, 0, 0, 0, 0),\\
    P_c &= (0,0,0,0,0,0,0, \sin a_1 \sin b_1).
\end{align}
We can find the tangent to this ellipse in the threshold point $P^* := P(\theta^*)$. Points, that lie on this tangent can be described as follows:
\begin{align}
    P_T(\gamma) &:= P^* + \gamma\left( \frac{\partial P}{\partial \theta} \right) \Big|_{\theta = \theta^*}=\\
    & P_0 + P_m (\cos \theta^* - \gamma \sin \theta^*) + P_c(\sin \theta^* + \gamma \cos \theta^*),
\end{align}
where $\gamma \in \R$. 

Now let us take $\gamma = \frac{\cos \theta^* - 1}{\sin \theta^*}$, then we obtain a point $P_E$ that lie on $8$ non-negativity facet and is a local point.
\begin{align}
    P_E = (1, \cos a_1, 1, \cos b_1, 1, \cos b_1, \cos a_1, 1 - |\cos a_1 - \cos b_1|).
\end{align}
It is easy to check, that $P_E$ lies inside the non-signaling set. The last thing to show is that it is local point. Let us show, that this point can be decomposed into three deterministic points. Let us assume, that $\cos a_1 \geq \cos b_1$ (the second case is analogical). Then this point can be decomposed as follows:
\begin{align}
    P_D^1 &:= (1,1,1,1,1,1,1,1),\\
    P_D^2 &:= (1,1,1,-1,1,-1,1,-1),\\
    P_D^3 &:= (1,-1,1,-1,1,-1,-1,1),\\
    \alpha_1 &:= \frac{1}{2}(1 + \cos b_1),\\
    \alpha_2 &:= \frac{1}{2}(\cos a_1 - \cos b_1),\\
    \alpha_3 &:= \frac{1}{2}(1 - \cos a_1),\\
    P_E &= \alpha_1 P_D^1 + \alpha_2 P_D^2 + \alpha_3 P_D^3.
\end{align}
Because of the fact that this is the local point, we can see, that all points with observables $a_0,b_0, a_1, b_1$ and $\sin \theta < \sin \theta^*$ can be decomposed into $P^*, P_E$ and some third point in the slice of the ellipse, for example $P_0$.

\bibliography{bibliography}

\begin{thebibliography}{25}%
\makeatletter
\providecommand \@ifxundefined [1]{%
 \@ifx{#1\undefined}
}%
\providecommand \@ifnum [1]{%
 \ifnum #1\expandafter \@firstoftwo
 \else \expandafter \@secondoftwo
 \fi
}%
\providecommand \@ifx [1]{%
 \ifx #1\expandafter \@firstoftwo
 \else \expandafter \@secondoftwo
 \fi
}%
\providecommand \natexlab [1]{#1}%
\providecommand \enquote  [1]{``#1''}%
\providecommand \bibnamefont  [1]{#1}%
\providecommand \bibfnamefont [1]{#1}%
\providecommand \citenamefont [1]{#1}%
\providecommand \href@noop [0]{\@secondoftwo}%
\providecommand \href [0]{\begingroup \@sanitize@url \@href}%
\providecommand \@href[1]{\@@startlink{#1}\@@href}%
\providecommand \@@href[1]{\endgroup#1\@@endlink}%
\providecommand \@sanitize@url [0]{\catcode `\\12\catcode `\$12\catcode
  `\&12\catcode `\#12\catcode `\^12\catcode `\_12\catcode `\%12\relax}%
\providecommand \@@startlink[1]{}%
\providecommand \@@endlink[0]{}%
\providecommand \url  [0]{\begingroup\@sanitize@url \@url }%
\providecommand \@url [1]{\endgroup\@href {#1}{\urlprefix }}%
\providecommand \urlprefix  [0]{URL }%
\providecommand \Eprint [0]{\href }%
\providecommand \doibase [0]{http://dx.doi.org/}%
\providecommand \selectlanguage [0]{\@gobble}%
\providecommand \bibinfo  [0]{\@secondoftwo}%
\providecommand \bibfield  [0]{\@secondoftwo}%
\providecommand \translation [1]{[#1]}%
\providecommand \BibitemOpen [0]{}%
\providecommand \bibitemStop [0]{}%
\providecommand \bibitemNoStop [0]{.\EOS\space}%
\providecommand \EOS [0]{\spacefactor3000\relax}%
\providecommand \BibitemShut  [1]{\csname bibitem#1\endcsname}%
\let\auto@bib@innerbib\@empty
\bibitem [{\citenamefont {Bell}(1964)}]{bellEinsteinPodolskyRosen1964}%
  \BibitemOpen
  \bibfield  {author} {\bibinfo {author} {\bibfnamefont {J.~S.}\ \bibnamefont
  {Bell}},\ }\href {\doibase 10.1103/PhysicsPhysiqueFizika.1.195} {\bibfield
  {journal} {\bibinfo  {journal} {Physics Physique Fizika}\ }\textbf {\bibinfo
  {volume} {1}},\ \bibinfo {pages} {195} (\bibinfo {year} {1964})}\BibitemShut
  {NoStop}%
\bibitem [{\citenamefont {Brunner}\ \emph {et~al.}(2014)\citenamefont
  {Brunner}, \citenamefont {Cavalcanti}, \citenamefont {Pironio}, \citenamefont
  {Scarani},\ and\ \citenamefont {Wehner}}]{brunnerBellNonlocality2014}%
  \BibitemOpen
  \bibfield  {author} {\bibinfo {author} {\bibfnamefont {N.}~\bibnamefont
  {Brunner}}, \bibinfo {author} {\bibfnamefont {D.}~\bibnamefont {Cavalcanti}},
  \bibinfo {author} {\bibfnamefont {S.}~\bibnamefont {Pironio}}, \bibinfo
  {author} {\bibfnamefont {V.}~\bibnamefont {Scarani}}, \ and\ \bibinfo
  {author} {\bibfnamefont {S.}~\bibnamefont {Wehner}},\ }\href {\doibase
  10.1103/RevModPhys.86.419} {\bibfield  {journal} {\bibinfo  {journal}
  {Reviews of Modern Physics}\ }\textbf {\bibinfo {volume} {86}},\ \bibinfo
  {pages} {419} (\bibinfo {year} {2014})},\ \Eprint
  {http://arxiv.org/abs/1303.2849} {arXiv:1303.2849 [quant-ph]} \BibitemShut
  {NoStop}%
\bibitem [{\citenamefont {{Herrero-Collantes}}\ and\ \citenamefont
  {{Garcia-Escartin}}(2017)}]{herrero-collantesQuantumRandomNumber2017}%
  \BibitemOpen
  \bibfield  {author} {\bibinfo {author} {\bibfnamefont {M.}~\bibnamefont
  {{Herrero-Collantes}}}\ and\ \bibinfo {author} {\bibfnamefont {J.~C.}\
  \bibnamefont {{Garcia-Escartin}}},\ }\href {\doibase
  10.1103/RevModPhys.89.015004} {\bibfield  {journal} {\bibinfo  {journal}
  {Reviews of Modern Physics}\ }\textbf {\bibinfo {volume} {89}},\ \bibinfo
  {pages} {015004} (\bibinfo {year} {2017})}\BibitemShut {NoStop}%
\bibitem [{\citenamefont {Pironio}\ \emph {et~al.}(2009)\citenamefont
  {Pironio}, \citenamefont {Ac{\'i}n}, \citenamefont {Brunner}, \citenamefont
  {Gisin}, \citenamefont {Massar},\ and\ \citenamefont
  {Scarani}}]{pironioDeviceindependentQuantumKey2009}%
  \BibitemOpen
  \bibfield  {author} {\bibinfo {author} {\bibfnamefont {S.}~\bibnamefont
  {Pironio}}, \bibinfo {author} {\bibfnamefont {A.}~\bibnamefont {Ac{\'i}n}},
  \bibinfo {author} {\bibfnamefont {N.}~\bibnamefont {Brunner}}, \bibinfo
  {author} {\bibfnamefont {N.}~\bibnamefont {Gisin}}, \bibinfo {author}
  {\bibfnamefont {S.}~\bibnamefont {Massar}}, \ and\ \bibinfo {author}
  {\bibfnamefont {V.}~\bibnamefont {Scarani}},\ }\href {\doibase
  10.1088/1367-2630/11/4/045021} {\bibfield  {journal} {\bibinfo  {journal}
  {New Journal of Physics}\ }\textbf {\bibinfo {volume} {11}},\ \bibinfo
  {pages} {045021} (\bibinfo {year} {2009})}\BibitemShut {NoStop}%
\bibitem [{\citenamefont {Boaron}\ \emph {et~al.}(2018)\citenamefont {Boaron},
  \citenamefont {Boso}, \citenamefont {Rusca}, \citenamefont {Vulliez},
  \citenamefont {Autebert}, \citenamefont {Caloz}, \citenamefont {Perrenoud},
  \citenamefont {Gras}, \citenamefont {Bussi{\`e}res}, \citenamefont {Li},
  \citenamefont {Nolan}, \citenamefont {Martin},\ and\ \citenamefont
  {Zbinden}}]{boaronSecureQuantumKey2018}%
  \BibitemOpen
  \bibfield  {author} {\bibinfo {author} {\bibfnamefont {A.}~\bibnamefont
  {Boaron}}, \bibinfo {author} {\bibfnamefont {G.}~\bibnamefont {Boso}},
  \bibinfo {author} {\bibfnamefont {D.}~\bibnamefont {Rusca}}, \bibinfo
  {author} {\bibfnamefont {C.}~\bibnamefont {Vulliez}}, \bibinfo {author}
  {\bibfnamefont {C.}~\bibnamefont {Autebert}}, \bibinfo {author}
  {\bibfnamefont {M.}~\bibnamefont {Caloz}}, \bibinfo {author} {\bibfnamefont
  {M.}~\bibnamefont {Perrenoud}}, \bibinfo {author} {\bibfnamefont
  {G.}~\bibnamefont {Gras}}, \bibinfo {author} {\bibfnamefont {F.}~\bibnamefont
  {Bussi{\`e}res}}, \bibinfo {author} {\bibfnamefont {M.-J.}\ \bibnamefont
  {Li}}, \bibinfo {author} {\bibfnamefont {D.}~\bibnamefont {Nolan}}, \bibinfo
  {author} {\bibfnamefont {A.}~\bibnamefont {Martin}}, \ and\ \bibinfo {author}
  {\bibfnamefont {H.}~\bibnamefont {Zbinden}},\ }\href {\doibase
  10.1103/PhysRevLett.121.190502} {\bibfield  {journal} {\bibinfo  {journal}
  {Physical Review Letters}\ }\textbf {\bibinfo {volume} {121}},\ \bibinfo
  {pages} {190502} (\bibinfo {year} {2018})},\ \Eprint
  {http://arxiv.org/abs/1807.03222} {arXiv:1807.03222 [quant-ph]} \BibitemShut
  {NoStop}%
\bibitem [{\citenamefont
  {Tsirel'son}(1987)}]{tsirelsonQuantumAnaloguesBell1987}%
  \BibitemOpen
  \bibfield  {author} {\bibinfo {author} {\bibfnamefont {B.~S.}\ \bibnamefont
  {Tsirel'son}},\ }\href {\doibase 10.1007/BF01663472} {\bibfield  {journal}
  {\bibinfo  {journal} {Journal of Soviet Mathematics}\ }\textbf {\bibinfo
  {volume} {36}},\ \bibinfo {pages} {557} (\bibinfo {year} {1987})}\BibitemShut
  {NoStop}%
\bibitem [{\citenamefont
  {Landau}(1988)}]{landauEmpiricalTwopointCorrelation1988}%
  \BibitemOpen
  \bibfield  {author} {\bibinfo {author} {\bibfnamefont {L.~J.}\ \bibnamefont
  {Landau}},\ }\href {\doibase 10.1007/BF00732549} {\bibfield  {journal}
  {\bibinfo  {journal} {Foundations of Physics}\ }\textbf {\bibinfo {volume}
  {18}},\ \bibinfo {pages} {449} (\bibinfo {year} {1988})}\BibitemShut
  {NoStop}%
\bibitem [{\citenamefont
  {Masanes}(2003)}]{masanesNecessarySufficientCondition2003}%
  \BibitemOpen
  \bibfield  {author} {\bibinfo {author} {\bibfnamefont {L.}~\bibnamefont
  {Masanes}},\ }\href@noop {} {\enquote {\bibinfo {title} {Necessary and
  sufficient condition for quantum-generated correlations},}\ } (\bibinfo
  {year} {2003}),\ \Eprint {http://arxiv.org/abs/quant-ph/0309137}
  {arXiv:quant-ph/0309137} \BibitemShut {NoStop}%
\bibitem [{\citenamefont {Le}\ \emph {et~al.}(2022)\citenamefont {Le},
  \citenamefont {Meroni}, \citenamefont {Sturmfels}, \citenamefont {Werner},\
  and\ \citenamefont {Ziegler}}]{leQuantumCorrelationsMinimal2022}%
  \BibitemOpen
  \bibfield  {author} {\bibinfo {author} {\bibfnamefont {T.~P.}\ \bibnamefont
  {Le}}, \bibinfo {author} {\bibfnamefont {C.}~\bibnamefont {Meroni}}, \bibinfo
  {author} {\bibfnamefont {B.}~\bibnamefont {Sturmfels}}, \bibinfo {author}
  {\bibfnamefont {R.~F.}\ \bibnamefont {Werner}}, \ and\ \bibinfo {author}
  {\bibfnamefont {T.}~\bibnamefont {Ziegler}},\ }\href {\doibase
  10.48550/arXiv.2111.06270} {\enquote {\bibinfo {title} {Quantum
  {{Correlations}} in the {{Minimal Scenario}}},}\ } (\bibinfo {year} {2022}),\
  \Eprint {http://arxiv.org/abs/2111.06270} {arXiv:2111.06270 [quant-ph]}
  \BibitemShut {NoStop}%
\bibitem [{\citenamefont {Yang}\ and\ \citenamefont
  {Navascues}(2013)}]{yangRobustSelfTesting2013}%
  \BibitemOpen
  \bibfield  {author} {\bibinfo {author} {\bibfnamefont {T.~H.}\ \bibnamefont
  {Yang}}\ and\ \bibinfo {author} {\bibfnamefont {M.}~\bibnamefont
  {Navascues}},\ }\href {\doibase 10.1103/PhysRevA.87.050102} {\bibfield
  {journal} {\bibinfo  {journal} {Physical Review A}\ }\textbf {\bibinfo
  {volume} {87}},\ \bibinfo {pages} {050102} (\bibinfo {year} {2013})},\
  \Eprint {http://arxiv.org/abs/1210.4409} {arXiv:1210.4409 [quant-ph]}
  \BibitemShut {NoStop}%
\bibitem [{\citenamefont {Bamps}\ and\ \citenamefont
  {Pironio}(2015)}]{bampsSumofsquaresDecompositionsFamily2015}%
  \BibitemOpen
  \bibfield  {author} {\bibinfo {author} {\bibfnamefont {C.}~\bibnamefont
  {Bamps}}\ and\ \bibinfo {author} {\bibfnamefont {S.}~\bibnamefont
  {Pironio}},\ }\href {\doibase 10.1103/PhysRevA.91.052111} {\bibfield
  {journal} {\bibinfo  {journal} {Physical Review A}\ }\textbf {\bibinfo
  {volume} {91}},\ \bibinfo {pages} {052111} (\bibinfo {year} {2015})},\
  \Eprint {http://arxiv.org/abs/1504.06960} {arXiv:1504.06960 [quant-ph]}
  \BibitemShut {NoStop}%
\bibitem [{\citenamefont {Wagner}\ \emph {et~al.}(2020)\citenamefont {Wagner},
  \citenamefont {Bancal}, \citenamefont {Sangouard},\ and\ \citenamefont
  {Sekatski}}]{wagnerDeviceindependentCharacterizationQuantum2020}%
  \BibitemOpen
  \bibfield  {author} {\bibinfo {author} {\bibfnamefont {S.}~\bibnamefont
  {Wagner}}, \bibinfo {author} {\bibfnamefont {J.-D.}\ \bibnamefont {Bancal}},
  \bibinfo {author} {\bibfnamefont {N.}~\bibnamefont {Sangouard}}, \ and\
  \bibinfo {author} {\bibfnamefont {P.}~\bibnamefont {Sekatski}},\ }\href
  {\doibase 10.22331/q-2020-03-19-243} {\bibfield  {journal} {\bibinfo
  {journal} {Quantum}\ }\textbf {\bibinfo {volume} {4}},\ \bibinfo {pages}
  {243} (\bibinfo {year} {2020})},\ \Eprint {http://arxiv.org/abs/1812.02628}
  {arXiv:1812.02628 [quant-ph]} \BibitemShut {NoStop}%
\bibitem [{\citenamefont {Hardy}(1992)}]{hardyQuantumMechanicsLocal1992}%
  \BibitemOpen
  \bibfield  {author} {\bibinfo {author} {\bibfnamefont {L.}~\bibnamefont
  {Hardy}},\ }\href {\doibase 10.1103/PhysRevLett.68.2981} {\bibfield
  {journal} {\bibinfo  {journal} {Physical Review Letters}\ }\textbf {\bibinfo
  {volume} {68}},\ \bibinfo {pages} {2981} (\bibinfo {year}
  {1992})}\BibitemShut {NoStop}%
\bibitem [{\citenamefont {Rai}\ \emph {et~al.}(2021)\citenamefont {Rai},
  \citenamefont {Pivoluska}, \citenamefont {Plesch}, \citenamefont {Sasmal},
  \citenamefont {Banik},\ and\ \citenamefont
  {Ghosh}}]{raiDeviceindependentBoundsCabello2021}%
  \BibitemOpen
  \bibfield  {author} {\bibinfo {author} {\bibfnamefont {A.}~\bibnamefont
  {Rai}}, \bibinfo {author} {\bibfnamefont {M.}~\bibnamefont {Pivoluska}},
  \bibinfo {author} {\bibfnamefont {M.}~\bibnamefont {Plesch}}, \bibinfo
  {author} {\bibfnamefont {S.}~\bibnamefont {Sasmal}}, \bibinfo {author}
  {\bibfnamefont {M.}~\bibnamefont {Banik}}, \ and\ \bibinfo {author}
  {\bibfnamefont {S.}~\bibnamefont {Ghosh}},\ }\href {\doibase
  10.1103/PhysRevA.103.062219} {\bibfield  {journal} {\bibinfo  {journal}
  {Physical Review A}\ }\textbf {\bibinfo {volume} {103}},\ \bibinfo {pages}
  {062219} (\bibinfo {year} {2021})},\ \Eprint
  {http://arxiv.org/abs/2103.09919} {arXiv:2103.09919 [quant-ph]} \BibitemShut
  {NoStop}%
\bibitem [{\citenamefont
  {Ishizaka}(2017)}]{ishizakaCryptographicQuantumBound2017}%
  \BibitemOpen
  \bibfield  {author} {\bibinfo {author} {\bibfnamefont {S.}~\bibnamefont
  {Ishizaka}},\ }\href {\doibase 10.1103/PhysRevA.95.022108} {\bibfield
  {journal} {\bibinfo  {journal} {Physical Review A}\ }\textbf {\bibinfo
  {volume} {95}},\ \bibinfo {pages} {022108} (\bibinfo {year}
  {2017})}\BibitemShut {NoStop}%
\bibitem [{\citenamefont
  {Ishizaka}(2018)}]{ishizakaNecessarySufficientCriterion2018}%
  \BibitemOpen
  \bibfield  {author} {\bibinfo {author} {\bibfnamefont {S.}~\bibnamefont
  {Ishizaka}},\ }\href {\doibase 10.1103/PhysRevA.97.050102} {\bibfield
  {journal} {\bibinfo  {journal} {Physical Review A}\ }\textbf {\bibinfo
  {volume} {97}},\ \bibinfo {pages} {050102} (\bibinfo {year}
  {2018})}\BibitemShut {NoStop}%
\bibitem [{\citenamefont
  {Ishizaka}(2020)}]{ishizakaGeometricalSelftestingPartially2020}%
  \BibitemOpen
  \bibfield  {author} {\bibinfo {author} {\bibfnamefont {S.}~\bibnamefont
  {Ishizaka}},\ }\href {\doibase 10.1088/1367-2630/ab6e49} {\bibfield
  {journal} {\bibinfo  {journal} {New Journal of Physics}\ }\textbf {\bibinfo
  {volume} {22}},\ \bibinfo {pages} {023022} (\bibinfo {year}
  {2020})}\BibitemShut {NoStop}%
\bibitem [{\citenamefont
  {Scarani}(2015)}]{scaraniDeviceindependentOutlookQuantum2015}%
  \BibitemOpen
  \bibfield  {author} {\bibinfo {author} {\bibfnamefont {V.}~\bibnamefont
  {Scarani}},\ }\href {\doibase 10.48550/arXiv.1303.3081} {\enquote {\bibinfo
  {title} {The device-independent outlook on quantum physics (lecture notes on
  the power of {{Bell}}'s theorem)},}\ } (\bibinfo {year} {2015}),\ \Eprint
  {http://arxiv.org/abs/1303.3081} {arXiv:1303.3081 [quant-ph]} \BibitemShut
  {NoStop}%
\bibitem [{\citenamefont {Goh}\ \emph {et~al.}(2018)\citenamefont {Goh},
  \citenamefont {Kaniewski}, \citenamefont {Wolfe}, \citenamefont
  {V{\'e}rtesi}, \citenamefont {Wu}, \citenamefont {Cai}, \citenamefont
  {Liang},\ and\ \citenamefont {Scarani}}]{gohGeometrySetQuantum2018}%
  \BibitemOpen
  \bibfield  {author} {\bibinfo {author} {\bibfnamefont {K.~T.}\ \bibnamefont
  {Goh}}, \bibinfo {author} {\bibfnamefont {J.}~\bibnamefont {Kaniewski}},
  \bibinfo {author} {\bibfnamefont {E.}~\bibnamefont {Wolfe}}, \bibinfo
  {author} {\bibfnamefont {T.}~\bibnamefont {V{\'e}rtesi}}, \bibinfo {author}
  {\bibfnamefont {X.}~\bibnamefont {Wu}}, \bibinfo {author} {\bibfnamefont
  {Y.}~\bibnamefont {Cai}}, \bibinfo {author} {\bibfnamefont {Y.-C.}\
  \bibnamefont {Liang}}, \ and\ \bibinfo {author} {\bibfnamefont
  {V.}~\bibnamefont {Scarani}},\ }\href {\doibase 10.1103/PhysRevA.97.022104}
  {\bibfield  {journal} {\bibinfo  {journal} {Physical Review A}\ }\textbf
  {\bibinfo {volume} {97}},\ \bibinfo {pages} {022104} (\bibinfo {year}
  {2018})},\ \Eprint {http://arxiv.org/abs/1710.05892} {arXiv:1710.05892
  [quant-ph]} \BibitemShut {NoStop}%
\bibitem [{\citenamefont
  {Masanes}(2005)}]{masanesExtremalQuantumCorrelations2005}%
  \BibitemOpen
  \bibfield  {author} {\bibinfo {author} {\bibfnamefont {L.}~\bibnamefont
  {Masanes}},\ }\href {\doibase 10.48550/arXiv.quant-ph/0512100} {\enquote
  {\bibinfo {title} {Extremal quantum correlations for {{N}} parties with two
  dichotomic observables per site},}\ } (\bibinfo {year} {2005}),\ \Eprint
  {http://arxiv.org/abs/quant-ph/0512100} {arXiv:quant-ph/0512100} \BibitemShut
  {NoStop}%
\bibitem [{\citenamefont {Acin}\ \emph {et~al.}(2012)\citenamefont {Acin},
  \citenamefont {Massar},\ and\ \citenamefont
  {Pironio}}]{acinRandomnessVsNon2012}%
  \BibitemOpen
  \bibfield  {author} {\bibinfo {author} {\bibfnamefont {A.}~\bibnamefont
  {Acin}}, \bibinfo {author} {\bibfnamefont {S.}~\bibnamefont {Massar}}, \ and\
  \bibinfo {author} {\bibfnamefont {S.}~\bibnamefont {Pironio}},\ }\href
  {\doibase 10.1103/PhysRevLett.108.100402} {\bibfield  {journal} {\bibinfo
  {journal} {Physical Review Letters}\ }\textbf {\bibinfo {volume} {108}},\
  \bibinfo {pages} {100402} (\bibinfo {year} {2012})},\ \Eprint
  {http://arxiv.org/abs/1107.2754} {arXiv:1107.2754 [quant-ph]} \BibitemShut
  {NoStop}%
\bibitem [{\citenamefont {Wolfe}\ and\ \citenamefont
  {Yelin}(2012)}]{wolfeNewQuantumBounds2012}%
  \BibitemOpen
  \bibfield  {author} {\bibinfo {author} {\bibfnamefont {E.}~\bibnamefont
  {Wolfe}}\ and\ \bibinfo {author} {\bibfnamefont {S.~F.}\ \bibnamefont
  {Yelin}},\ }\href {\doibase 10.1103/PhysRevA.86.012123} {\bibfield  {journal}
  {\bibinfo  {journal} {Physical Review A}\ }\textbf {\bibinfo {volume} {86}},\
  \bibinfo {pages} {012123} (\bibinfo {year} {2012})},\ \Eprint
  {http://arxiv.org/abs/1106.2169} {arXiv:1106.2169 [math-ph,
  physics:quant-ph]} \BibitemShut {NoStop}%
\bibitem [{\citenamefont {Chen}\ \emph {et~al.}(2022)\citenamefont {Chen},
  \citenamefont {Tabia}, \citenamefont {Jebarathinam}, \citenamefont {Mal},
  \citenamefont {Wu},\ and\ \citenamefont
  {Liang}}]{chenQuantumCorrelationsNosignaling2022}%
  \BibitemOpen
  \bibfield  {author} {\bibinfo {author} {\bibfnamefont {K.-S.}\ \bibnamefont
  {Chen}}, \bibinfo {author} {\bibfnamefont {G.~N.~M.}\ \bibnamefont {Tabia}},
  \bibinfo {author} {\bibfnamefont {C.}~\bibnamefont {Jebarathinam}}, \bibinfo
  {author} {\bibfnamefont {S.}~\bibnamefont {Mal}}, \bibinfo {author}
  {\bibfnamefont {J.-Y.}\ \bibnamefont {Wu}}, \ and\ \bibinfo {author}
  {\bibfnamefont {Y.-C.}\ \bibnamefont {Liang}},\ }\href@noop {} {\enquote
  {\bibinfo {title} {Quantum correlations on the no-signaling boundary:
  Self-testing and more},}\ } (\bibinfo {year} {2022}),\ \Eprint
  {http://arxiv.org/abs/2207.13850} {arXiv:2207.13850 [quant-ph]} \BibitemShut
  {NoStop}%
\bibitem [{\citenamefont {Rai}\ \emph {et~al.}(2022)\citenamefont {Rai},
  \citenamefont {Pivoluska}, \citenamefont {Sasmal}, \citenamefont {Banik},
  \citenamefont {Ghosh},\ and\ \citenamefont
  {Plesch}}]{raiSelftestingQuantumStates2022}%
  \BibitemOpen
  \bibfield  {author} {\bibinfo {author} {\bibfnamefont {A.}~\bibnamefont
  {Rai}}, \bibinfo {author} {\bibfnamefont {M.}~\bibnamefont {Pivoluska}},
  \bibinfo {author} {\bibfnamefont {S.}~\bibnamefont {Sasmal}}, \bibinfo
  {author} {\bibfnamefont {M.}~\bibnamefont {Banik}}, \bibinfo {author}
  {\bibfnamefont {S.}~\bibnamefont {Ghosh}}, \ and\ \bibinfo {author}
  {\bibfnamefont {M.}~\bibnamefont {Plesch}},\ }\href {\doibase
  10.1103/PhysRevA.105.052227} {\bibfield  {journal} {\bibinfo  {journal}
  {Physical Review A}\ }\textbf {\bibinfo {volume} {105}},\ \bibinfo {pages}
  {052227} (\bibinfo {year} {2022})},\ \Eprint
  {http://arxiv.org/abs/2112.06595} {arXiv:2112.06595 [quant-ph]} \BibitemShut
  {NoStop}%
\bibitem [{\citenamefont {Basu}\ \emph {et~al.}(2017)\citenamefont {Basu},
  \citenamefont {Cornuejols},\ and\ \citenamefont
  {Zambelli}}]{basuConvexSetsMinimal2017}%
  \BibitemOpen
  \bibfield  {author} {\bibinfo {author} {\bibfnamefont {A.}~\bibnamefont
  {Basu}}, \bibinfo {author} {\bibfnamefont {G.}~\bibnamefont {Cornuejols}}, \
  and\ \bibinfo {author} {\bibfnamefont {G.}~\bibnamefont {Zambelli}},\ }\href
  {\doibase 10.48550/arXiv.1701.06550} {\enquote {\bibinfo {title} {Convex
  {{Sets}} and {{Minimal Sublinear Functions}}},}\ } (\bibinfo {year} {2017}),\
  \Eprint {http://arxiv.org/abs/1701.06550} {arXiv:1701.06550 [math]}
  \BibitemShut {NoStop}%
\end{thebibliography}%
\end{document}